\documentclass[aps,twocolumn,preprintnumbers,amsmath,amssymb]{revtex4-2}

\usepackage{graphicx}
\usepackage{dcolumn}
\usepackage{bm}

\usepackage{color}
\usepackage[normalem]{ulem}
\usepackage{hyperref}

\usepackage{appendix}

\usepackage[table,xcdraw]{xcolor}

\begin{document}

\title{Non-classical nucleation pathways in stacking-disordered crystals}
\author{Fabio Leoni}
\email{\text{fabio.leoni@uniroma1.it}}
\affiliation{Department of Physics, Sapienza University of Rome, P.le Aldo Moro 5, 00185 Rome, Italy}
\author{John Russo}
\email{\text{john.russo@uniroma1.it}}
\affiliation{Department of Physics, Sapienza University of Rome, P.le Aldo Moro 5, 00185 Rome, Italy}
\affiliation{School of Mathematics, University of Bristol, Bristol BS8 1UG, United Kingdom}

\date{\today}

\begin{abstract}
The nucleation of crystals from the liquid melt is often characterized by a competition between different crystalline structures or polymorphs, and can result in nuclei with heterogeneous compositions. These mixed-phase nuclei can display nontrivial spatial arrangements, such as layered and onion-like structures, whose composition varies according to the radial distance, and which so far have been explained on the basis of bulk and surface free-energy differences between the competing phases.
Here we extend the generality of these non-classical nucleation processes, showing that layered and onion-like structures can emerge solely based on structural fluctuations even in absence of free-energy differences.
We consider two examples of competing crystalline structures, \emph{hcp} and \emph{fcc} forming in hard spheres, relevant for repulsive colloids and dense liquids, and the cubic and hexagonal diamond forming in water, relevant also for other group 14 elements such as carbon and silicon.
We introduce a novel structural order parameter that combined with a neural network classification scheme allows us to study the properties of the growing nucleus from the early stages of nucleation.
We find that small nuclei have distinct size fluctuations and compositions from the nuclei that emerge from the growth stage. The transition between these two regimes is characterized by the formation of onion-like structures, in which the composition changes with the distance from the center of the nucleus, similarly to what seen in two-step nucleation process.
\end{abstract}


\maketitle

\section{Introduction}

Nucleation is a discontinuous transition in which clusters of molecules self assemble due to fluctuations that are very localized in space and time to form a growing nucleus. It is a crucial phenomenon in many fields of natural science \cite{moore2011structural,bartels2012,sosso2016}, going from planetary- to nano-scale.
During the nucleation process of many materials, including several metals, minerals and polymers, different crystalline phases, called polymorphs, can nucleate.
The structure of the growing nucleus in such materials can depend on many, eventually size-dependent \cite{leoni2019,molinero2017nature}, effects, such as energy and entropy competition, or frustration.
Understanding the selection mechanism of polymorphs is fundamental to predict the structure of the growing nucleus, with applications ranging from Earth's weather and climate forecast, especially in relation to the formation of nanometer-sized ice crystallites in clouds \cite{kaufman2002,murray2005,sastry2005,sassen2005,herbert2015,shaw2005}, to the pharmaceutic industry, where the physical and chemical properties of the drug molecules can change with the eventual crystallization of unwanted polymorph forms~\cite{lee2011}.
For example, the molecule for aspirin (acetylsalicylic acid), one of the most widely consumed medications, has two polytypic crystalline forms~\cite{vishweshwar2005predictably}.

Here we study the nucleation of polytypes, a specific type of polymorphs where the crystalline structures have the same projection along a specific direction, and only differ in the way the planes perpendicular to that direction are stacked onto each other. Some of the most common crystalline structures formed in metals are polytypic, notably the \emph{hcp} (hexagonal close packed) and \emph{fcc} (face centered cubic) crystalline structures, and the hexagonal and cubic diamond forms.

We consider the formation of polytypes in two important systems: the hard-sphere (HS) model and the coarse-grained mW model of water. 
They are representative of a wide class of materials, repulsive colloids and dense liquids \cite{berthier2009,bacher2014} for HS, and tetrahedrally-bonded materials (like water and group 14 elements such as carbon and silicon) for mW \cite{molinero2009,moore2011structural}. They crystallize in two different polytypes: either  \emph{fcc} or \emph{hcp} for HS, and either cubic ice (I$_c$) or hexagonal ice (I$_h$) for water.
Importantly, in both cases the difference in all thermodynamic relevant quantities (such as free-energy difference, nucleation barrier, and solid/melt surface tension) between the competing polytypes are negligibly small (within $10^{-3}\,k_BT$ per particle for all cases)~\cite{pronk1999,molinero2017nature,malkin2015stacking,moore2011structural,zaragoza2015,cheng2018theoretical,quigley}.
For example, in the mW water model \cite{molinero2009} the stacking fault between the ice I$_c$ and I$_h$ has been estimated as low as $ 0.16\pm 0.05$~mJ~m$^{-2}$ at $T=218$~K~\cite{leoni2019}.
In this way, the nucleation mechanisms for both systems are determined not by bulk free energy properties, or by details of their interactions, but by general principles, which we aim at elucidating in the present work.

One of the main difficulties in studying polymorph composition is assigning the local environments surrounding a particle to a particular phase, distinguishing between amorphous (liquid) structures and crystalline ones~\cite{tanaka2019revealing}.
Several methods for local structure identification have been developed so far.
Contrary to common belief, the method employed to classify a single particle as belonging to a specific polymorph can sensibly alter the measured composition of the nucleus \cite{allahyarov2015crystallization,prestipino2018}.
In the present work we compare some of the more representative methods found in the literature and introduce new methods which allow us to find fundamental properties of the nucleus growing during the early stage of homogeneous nucleation, and in particular to find evidences of a two-step nucleation (TSN) pathway.

Departures from single-step nucleation have been recently observed in the nucleation of polymorphs \cite{kashchiev2005kinetics,erdemir2009nucleation,vekilov2010two,schilling2010precursor,haxton2015crystallization,russo2016nonclassical,sosso2016crystal} in systems as different as hard particle fluids~\cite{lee2019entropic}, colloids \cite{tan2014}, salt solutions~\cite{jiang2019nucleation}, and calcium carbonate formation \cite{gebauer2008,pouget2009,sear2012}.

Two-step nucleation mechanisms  involve one disordered phase (the melt) and (at least) two ordered crystalline phases. 
Two-step nucleation often produces layered structures, where the composition changes radially within the nucleus, in such a way that the most stable polymorph form is closer to the centre, and is ``wetted'' by the metastable form on the surface. This structure, often referred to as the \emph{onion} structure, is one of the hallmarks of two-step nucleation pathways, that have been theoretically predicted via classical nucleation theory (CNT)~\cite{kashchiev2005kinetics,kashchiev2020classical}, density functional theory ~\cite{toth2011amorphous,barros2013liquid,santra2013nucleation,lutsko2019crystals}, phase-field models~\cite{tang2017competitive}, two-dimensional lattice models~\cite{james2019phase}, and molecular simulations~\cite{desgranges2019can}.

These two-step nucleation pathways are generally explained via free energy differences between the two crystalline phases, and in particular with the different surface free energy of the crystals with respect to the melt~\cite{james2019phase}.
Instead, in the systems under consideration the polytypes in competition have the same bulk free energy properties, and classical theory would predict in this case a homogeneous composition of the nuclei. We will observe that, due to finite-size fluctuation effects, onion-like structures are formed also under these conditions and that well separated free energy channels, corresponding to the competing polymorphs, can be distinguished, extending the phenomenology of structured nuclei to this large family of crystals.

The outline of the article is the following: in Sec.~\ref{sec:PI} we describe the methods for local structure identification employed in the present work to study the properties of nuclei forming during the homogeneous nucleation of HS and mW water. In Sec.~\ref{sec:MS} we describe the model systems we simulated, HS and mW water. In Sec.~\ref{sec:Results} we compare the properties of nuclei as obtained by using the methods described in Sec.~\ref{sec:PI}. Finally, in Sec.~\ref{sec:Conclusions} we present concluding remarks.

\section{Particle identification}
\label{sec:PI}

\subsection{Order parameters}

Common widespread methods used in the literature to identify local structures usually employ one- or two-dimensional order parameters (OP) maps, which involve the comparison of the local environment of a particle with different reference structures. 
For this reason, thresholds are usually introduced to establish which reference structure the particle under investigation belongs to. Steinhardt or bond orientational order (BOO) parameters in their averaged form $\bar{q}_l$ \cite{steinhardt1983} (see Appendix \ref{app:A})
are the standard choice as OP, where four-fold ($l=4$) and six-fold ($l=6$) are often the only symmetries considered.
Other methods involve the study of topological properties of the bond network, such as the Common neighbor Analysis (CNA) method.For water-like systems, the CNA method considers also second-nearest neighbours, and is named Ext-CNA.
In Appendix \ref{app:A} we describe some of the most representative low-dimensional OP employed in the present study for local structure identification (for a comprehensive review on common OP see Ref.~\cite{tanaka2019revealing}).We also present some tests aimed at determining the accuracy of the different methods in controlled situations (Appendix~\ref{sec:benchmark}).
Since previous \emph{low-dimensional} OP have produced different results when applied to the model systems studied here~\cite{filion2010,taffs2013,molinero2017nature,chan2019,niu2019,allahyarov2015crystallization,martelli2018searching,prestipino2018}, we consider a high-dimensional OP based on 30 BOO (see Appendix \ref{app:A}). In the following we drop the number 30 and use only the acronym BOO to refer to this method. 
However, the degeneracy of OP like BOO, for which the same OP value can correspond to different local environments \cite{lazar2015}, could result, in some specific application, in a suboptimal performance due to misidentification.
For example, the use of Steinhardt OP, especially of those related to the spherical harmonic with angular momentum $l=3$ and $m=2$ ($Y_{32}$), which is the only one with tetrahedral geometry, to distinguish I$_c$ from I$_h$ in water has been already questioned in previous works \cite{brukhno2008challenges}.
In order to resolve also the issue related to the degeneracy of the OP, in the following section we introduce a novel \emph{lossless} order parameter for the characterization of local environments.

\subsection{Local inter-distance (LID)}

Here we introduce a novel order parameter for the characterization of local environments that is built according to the following two principles. Firstly, the OP is \emph{high-dimensional}: increasing the dimensionality of the order parameter space allows to easily increase the separation between the different populations of local environments we want to discriminate between. Secondly, the OP is \emph{lossless}: with this we mean that no information is lost by going from the real-space coordinates of the particles in the environment under consideration to its order-parameter representation; in other words, from the OP it is possible to reconstruct the original positions of the particles, except for translations, rotations, or particle-index permutations.
Indeed, this method is based on the distances between all possible pairs obtained from a particle and its neighbors, and the problem to establish whether to the set of all possible inter-distances between a number of points corresponds only one points configuration dates back to the problem of the uniqueness in the x-ray analysis of crystal structures \cite{patterson1944}, in which case only very few specific exceptions are known.

The new order parameter is inspired by the permutation invariant vector of Ref. \cite{gallet2013,pipolo2017} and the Deep Potential Molecular Dynamics method of Ref.~\cite{zhang2018}, and is constructed in the following way: for each particle $i$ we make a list of its first ($f^j_i$) and second ($s^k_i$) nearest neighbors, with $j=1\cdots N$ and $k=1\cdots M$, where $N$ and $M$ is the number of first and second nearest neighbors respectively. 
We then compute all the $(N+M+1)(N+M)/2$ possible distances $d_{pq}=|\vec{r}_p-\vec{r}_q|$ between particle $p$ and particle $q$ with $p,q=1,2,....,N+M+1$ and $p\neq q$ and subdivide them in the following groups.
For HS we group the $d_{pq}$ in 5 categories: $(i,f_i^j)$ (12 terms), $(i,s_i^k)$ (6 terms), $(f_i^{j'},f_i^{j''})$ (66 terms), $(s_i^{k'},s_i^{k''})$ (15 terms), $(f_i^j,s_i^k)$ (72 terms).
In mW water we group the $d_{pq}$ in 6 categories where now $f_i^j$ and $s_i^k$ are the first and second energetic neighbors of particle $i$.
The 6 categories are: $(i,f_i^j)$ (4 terms), $(f_i^{j'},f_i^{j''})$ (6 terms), $(f_i^j,s_i^{k_j})$ (12 terms), $(i,s_i^k)$ (12 terms), $(s_i^{k'},s_i^{k''})$ (66 terms), and $(f_i^j,s_i^k)$ (36 terms) where $s_i^{k_j}$ is a second neighbor of particle $i$ which is also first neighbor of particle $f_i^j$.
The number of terms in each category is obtained by considering $N=12$ and $M=6$ for HS, while $N=4$ and $M=12$ for mW water. These values for $N$ and $M$ are related to the number of first and second neighbors in the crystalline structures forming in these models.

The distances in each group are then sorted in ascending order. This makes the OP invariant under particle-index permutations.
Since in the NN we use the Sigmoid as activation function (see Sec.\ref{sec:NN}), which works better with inputs between $-1$ and $1$, we normalize the grouped and sorted distances $d_{pq}^{g,s}$ for the average local environment radius $r_0$ (considering the first neighbors shell for HS and up to the second shell for mW water), and subtract from it the total normalized inter-distances $\langle d_{pq}^{g,s}/r_0\rangle_{out}$ (considering all outputs of the NN).
Finally, the order parameter we introduce here, which we name LID (local inter-distance), is the vector obtained from the union of all the groups: $d_{pq}^{g,s,n}=d_{pq}^{g,s}/r_0-\langle d_{pq}^{g,s}/r_0\rangle_{out}$. 

To emphasize the advantages of LID we will compare its results with the ones obtained via either a low-dimensional method, i.e. Common neighbor Analysis (CNA), or via a high-dimensional (but not lossless~\cite{brukhno2008challenges}) order parameter constructed as an array of 30 different bond orientational order (BOO) parameters (built from spherical harmonic invariants of order up to $l=12$, see Appendix~\ref{app:A}).

\subsection{Neural Networks classification scheme}
\label{sec:NN}

To partition a multidimensional OP space in different volumes, each one associated with the local environment of a crystalline structure or liquid phase, we use artificial neural networks (NN).
In condensed matter NN have been used for potential energy surface calculations \cite{behler2007,zhang2018},
to construct accurate molecular force fields \cite{chmiela2017}, 
to improve potential energy of coarse grained models for water \cite{chan2019}, or for identification and classification of local ordered or disordered structures using supervised \cite{geiger2013,bapst2020,martelli2020} and unsupervised \cite{spellings2018,reinhart2017,reinhart2017b,boattini2018,boattini2019,adorf2019analysis} methods.
Ideally, unsupervised learning allows to cluster high-dimensional OP space into sets corresponding to different structures before they have been identified \cite{spellings2018,reinhart2017,reinhart2017b,boattini2019}. 
If all possible structures present in the system are known a priori, supervised learning is a powerful method to identify local structures not requiring any threshold chosen ad hoc and being less sensitive to hyper-parameters.
We choose here supervised training, in which the NN is first trained against sample configurations of the phases we are interested in identifying. For the training we use bulk configurations prepared at coexistence conditions, where thermal fluctuations in the solid phases are maximized. For the HS system this corresponds to preparing bulk \emph{fcc}, \emph{hcp}, \emph{bcc}, and fluid configurations at pressure $P=11.54$ (in conventional reduced units) \cite{noya2008}, and running event driven molecular dynamics simulations, and using each local environment as a training sample. 
In detail, the training set for HS is obtained by 10 different realization of \emph{fcc}, \emph{hcp} and \emph{bcc} crystals at the melting point ($\phi=0.545$), and 20 different realization of the liquid phase at the freezing point ($\phi=0.494$), all composed of $N\sim 10000$ particles.
The training set for mW water is obtained by running Monte Carlo simulations at ambient pressure of 10 different realization of I$_c$, I$_h$ at the melting temperature $T_m=275K$, and of ice~0 at its melting temperature $T_m=244K$ (being metastable it has a lower melting temperature), and 20 different realization of the liquid phase at the melting temperature $T_m=275K$, all composed of $N=5376$ particles.
We notice here that we don't train the NN against surfaces, as these would require an external criteria in order to be defined (such as the Gibbs dividing surface), and we are only interested in bulk-like local environments.
We choose a single-layer feed-forward network topology.
As descriptors we consider the 30-dimensional (both for HS and mW water) BOO OP, and the 171-dimensional (for HS) and the 136-dimensional (for mW water) LID OP.
The hidden layer (HL) for BOO (both for HS and mW water) is composed of 8 nodes. We obtained the same performance varying the number of nodes in the HL from 4 to 20, indicating that the network is quite robust.
The HL in LID is composed of 30 nodes (both for HS and mW water). Also in this case we observe the same performance of the network for a wide range of nodes in the HL.
We initialize the weights following the Xavier method \cite{glorot2010}, 
consisting to set random weights from a normal distribution with zero mean and variance equal to 2 divided by the sum of the number of nodes in the input layer and the output layer.
We consider the Sigmoid, or logistic function, as activation function for both IL-HL and HL-OL, where IL and OL are the input and output layer, respectively.
The OL is composed of 4 nodes, which correspond to the 4 possible phases identified during the homogeneous nucleation of HS and mW water at the thermodynamic conditions considered here.
As error or loss function we take the overall mean square error between the actual and the target output. 
We minimize the error using the stochastic gradient descent (for a critical discussion see~\cite{zhang2018energy}) and update the weights following the back-propagation approach \cite{bishop1995}.
The performance of the NN is higher than 98\% in all cases. The absence of overfitting is verified by obtaining the same performance considering both the test and the training set.
For all cases we set the learning rate to $\alpha=0.01$, while the number of epochs is 50 for BOO and 100 for LID for both HS and mW water.

\begin{figure*}[!t] 
	\begin{center}
		\includegraphics[width=17cm]{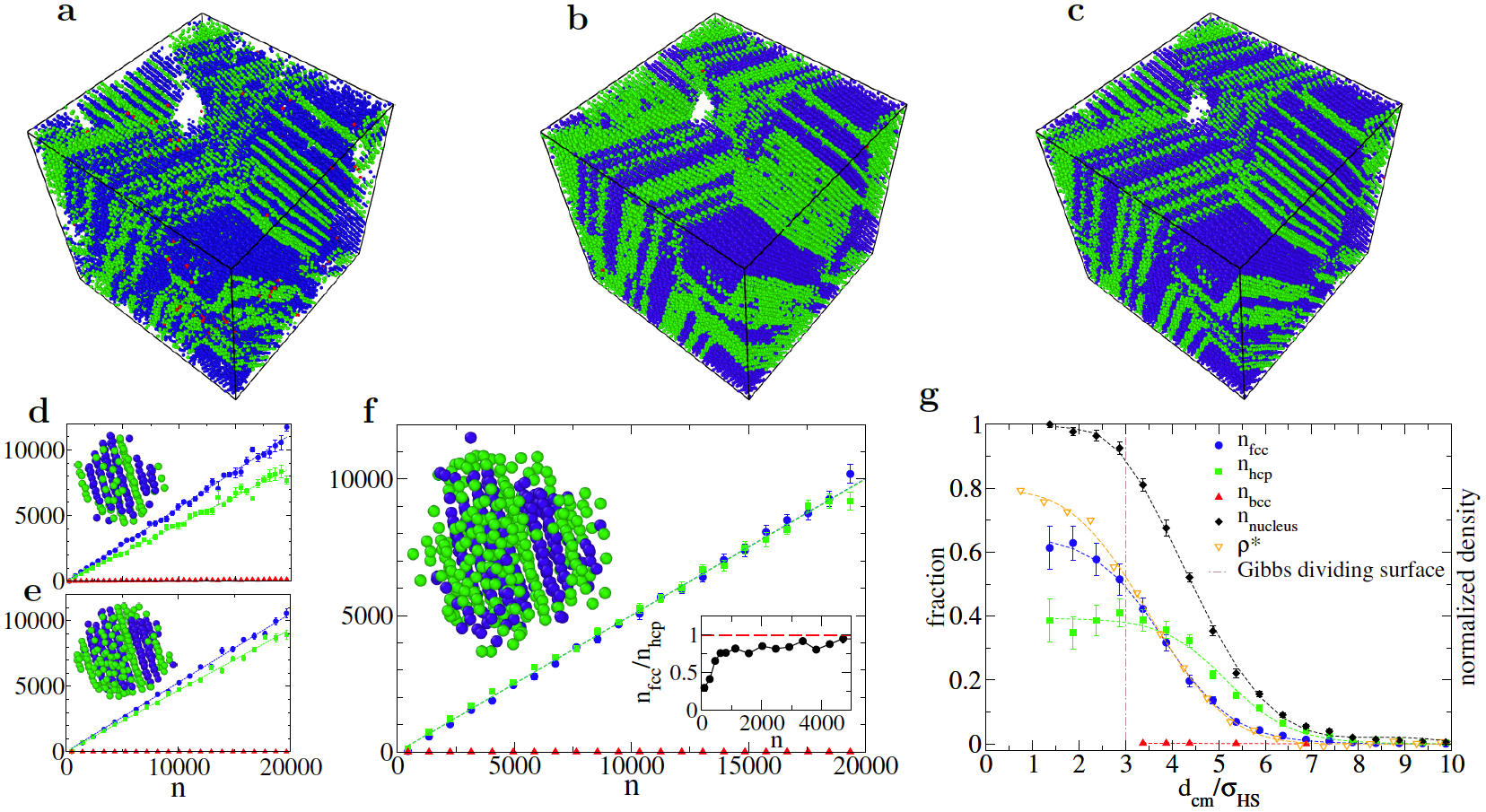}
                \caption{\label{fig:nc_HS} Homogeneous nucleation of hard spheres.
    (a), (b), (c) Snapshots of crytalline grains spanning the simulation box. Particle structures from the same configuration have been identified using the following methods: CNA (a), BOO (b), and LID (c).
    In all panels the colors associated to \emph{fcc}, \emph{hcp}, and \emph{bcc} structures are blue, green, and red, respectively.
    We show average composition of the main cluster as identified by CNA (d), BOO (e), and LID (f).
    Insets in (d), (e) and (f) show a typical nucleus composed of 188, 398 and 502 particles, respectively. The inset to the right in (f) shows the average ratio $r=n_{fcc}/n_{hcp}$ between the number of particles composing the nucleus in the \emph{fcc} and \emph{hcp} phase using LID for local structure identification.
    (g) Average radial fractional composition of the main cluster (for clusters of size $500\leq n\leq 550$) as identified by LID. $d_\text{cm}$ is the distance from the center of mass of the cluster and $\sigma_{HS}$ the hard sphere diameter. Dashed fitting lines are a guide for the eyes. Snapshots obtained by using \emph{Ovito} \cite{stukowski2010}.}
	\end{center}
\end{figure*}

\subsection{Nucleus identification}
\label{sec:NC}

After all particles in the system have been classified as belonging to a specific phase, in order to identify clusters of solid particles we use the same method employed in Ref.~\cite{wolde1996}: two solid particles are considered to belong to the same cluster if their distance is smaller than the value of the first minimum of the radial distribution function of the liquid (which results to be  $\sim1.5\sigma_{HS}$ in HS, and $\sim1.5\sigma_{mW}$ in mW water). 
After a solid particle has been added to a cluster, the enumeration needed to distinguish the different clusters is obtained by using the Hoshen-Kopelman algorithm \cite{hoshen1976}.

Other methods used to identify neighbors are the Voronoi construction, which is parameter free, but computationally expensive and sensitive to thermal fluctuations \cite{meel2012,saija2001}, 
and the solid-based nearest-neighbor (SANN) algorithm \cite{meel2012}, which is parameter free and more robust against thermal fluctuations respect to the Voronoi construction, but occasionally can include second shell neighbors in the first shell neighbors \cite{boattini2018}.

\section{Model systems}
\label{sec:MS}

\subsection{Homogeneous nucleation of hard spheres}

Here we consider non-overlapping hard spheres, the reference model for systems with excluded volume interactions~\cite{hansen2013theory}.
We perform event driven molecular simulation of N=100000 hard spheres at constant volume $V$ using the open-source event-driven particle simulator DynamO \cite{bannerman2011}.
The phase diagram of hard spheres is a function of $\phi=Nv/V$, which is the fraction of the box volume V covered by the N spheres, each sphere having volume $v=(\pi/6)\sigma_{HS}$. $\sigma_{HS}$ is the sphere diameter.
We consider 100 different trajectories simulating different initial configuration of supersaturated fluids at volume fraction $\phi=0.535$, between the freezing $\phi=0.494$ and the melting $\phi=0.545$ value. 
Each configuration of supersaturated fluid is obtained using a Monte Carlo method which moves, consisting on expansion of spheres diameter, are rejected if at least two spheres volume overlap. 
$\phi=0.535$ is close enough to the melting value such for the supersaturated fluid to nucleate easily, but far enough such to avoid multiple critical nuclei growing and eventually merging together. Indeed, in all the 100 different trajectories simulated, we always observe one critical nucleus growing within the maximum number of collisions simulated, which is $10^{10}$, i.e. an average of $2\times10^{5}$ collisions per particle. 
We also performed simulations for $\phi=0.54$ and observed multiple nuclei growing and merging during the nucleation process.

\subsection{Homogeneous nucleation of mW water}

The mW model of water is a popular coarse grained representation of water, where the molecule is replaced by a single site having both two-body and three-body interactions~\cite{molinero2009,moore2011structural}.
We perform Monte Carlo simulations of N=4000 mW particles in the NPT ensemble at pressure P=0~Pa and temperature T=204~K.
At these thermodynamic conditions the mW water model spontaneously nucleates within the maximum time simulated.
We consider 100 different trajectories simulating different initial configuration of supercooled fluid. 
Each supercooled fluid configuration is obtained using the same Monte Carlo method employed to get supersaturated HS fluid.

A system with directional tetrahedral interaction has the potential to offer additional insights on nucleation pathways, as, in principle, it can involve many polymorphic structures~\cite{engel2018mapping}. We focus here on the stable ice $I$ polytypes (the cubic form I$_c$, and the hexagonal form I$_h$), and on the metastable ice~0 structure~\cite{russo2014new,bordonskiy2017}. We choose this polymorph as it is currently the only known structure to satisfy all the following criteria: it has the lowest free energy outside the stable cubic and hexagonal (ice~$I$) structures~\cite{slater2014crystal}; it is the simplest structure that can be built by deformation of the diamond crystal while preserving to a large degree a highly regular fourfold coordination for the sites~\cite{mujica2015low}; it can stack coherently (without breaking of bonds between grains) with the diamond crystal~\cite{leoni2019}. These structures have never been observed as fully formed crystals, and instead we focus on clusters of molecules whose nearest-neighbor environment is close to those found in the bulk ice~0 crystal. It has been recently shown that these clusters have a lower energy than their stable ice $I$ counterparts up to cluster sizes of around 40 water molecules~\cite{leoni2019}.

\section{Results} 
\label{sec:Results}

\subsection{Hard spheres}

\subsubsection{Nucleus composition}

In Fig.~\ref{fig:nc_HS} we show the results from the homogeneous nucleation of the HS system obtained from 100 independent event-driven molecular dynamics~\cite{bannerman2011} trajectories of $10^5$ particles at the volume fraction $\phi=0.535$. The snapshots in panels \textit{a}, \textit{b}, \textit{c} compare the same simulation configuration of large-scale grains, which are colored according to the classification output of CNA, BOO, and LID order parameters respectively. The color indicates the detected phase: blue, green, red respectively for the \emph{fcc}, \emph{hcp}, and \emph{bcc} local environments. Already from a quick visual inspection, we see that both the CNA and BOO methods have a lower resolution of grains respect to LID whenever there is a high degree of \emph{hcp} and \emph{fcc} stacking. On a quantitative level, panels \textit{d}, \textit{e}, \textit{f} report, for the same order parameters, the average fraction of the different polymorphs within the largest nucleus as a function of the nucleus size $n$. All methods do not detect any \emph{bcc} in the nucleus, as was already found in Ref.~\cite{auer2001prediction}. Both the \emph{fcc} and \emph{hcp} fractions instead grow linearly (volume-scaling) with $n$. If we define the ratio $r=n_{fcc}/n_{hcp}$ (where $n_{fcc}$ and $n_{hcp}$ are the number of particles in the \emph{fcc} and \emph{hcp} phase, respectively) we see that the low-dimensional method CNA gives a value ($r=1.31 \pm 0.05$) that is incompatible with both multidimensional methods: $r=1.07 \pm 0.05$ for BOO and $r=1.00 \pm 0.05$ for LID. A ratio $r\sim 1$ is indeed expected during the growth stage given the low free energy difference between the \emph{fcc} and \emph{hcp} phases and the fact that the crystals are polytypes, i.e. they can stack onto each other with considerable entropy gain~\cite{molinero2017nature}. Both multidimensional methods agree within the error. The snapshots in panels \textit{d}, \textit{e}, \textit{f} show a nucleus identified by the different order parameters from the same configuration. We note that the number of particles identified as crystalline varies considerably depending on the method: 188 particles for CNA, 398 for BOO, and 502 for LID. The multidimensional methods that use the NN detect larger nuclei as they have been trained with configurations of crystal structures at melting, thus including as much thermal fluctuations as possible without breaking the crystal order. LID, as we will confirm below for the mW model, is particularly effective even for distorted local environments.

In Fig.~\textit{1g} we focus on the LID method and show both the composition (full symbols) and density (open symbols) profiles, as a function of the distance from the centre of mass of the nucleus, averaged over nuclei of size $500\leq n\leq550$ . This size was chosen to be well above the critical nucleus size: from a mean-first passage time of the nucleating trajectories (see section \ref{sec:critical_nucleus} for a theoretical description in the case of mW water) we estimate the critical size to be $n_c\sim 180$ for the LID order parameter, meaning that the profiles in Fig.~\textit{1g} are for nuclei which are three times this size. The figure reveals two important characteristics of two-step nucleation pathways. The first is the decoupling between the density and structural order fields. The open symbols represent the normalized density $\rho^*=(\rho-\rho_{f})/(\rho_{x}-\rho_{f})$, such that the values $0$ and $1$ are assigned respectively to the bulk density of the fluid and crystal phases;  $\rho=1/\langle v\rangle$ and $\langle v\rangle$ is the average specific volume computed via a Voronoi tesselation. As is seen here, and contrary to CNT assumptions, the nucleus reaches only about $80\%$ of its bulk density close to the centre. Recently, for HS it has been confirmed that using CNT in combination with bulk quantities yields inaccurate results in the description of nucleation \cite{richard2018crystallization}.
The second characteristic is the difference in profiles for the \emph{fcc} (blue symbols) and \emph{hcp} (green symbols) polytypes. While the \emph{fcc} phase is found more abundantly near the centre of mass of the nucleus, \emph{hcp} has a relative higher concentration towards the surface with the fluid. This is the \emph{onion} structure mentioned before. In the next section we examine in more detail the nucleation pathway of these structures.

\subsubsection{Onion-like structures}
\label{sec:small_HS}


\begin{figure}[!t] 
	\begin{center}
		\includegraphics[width=8.5cm]{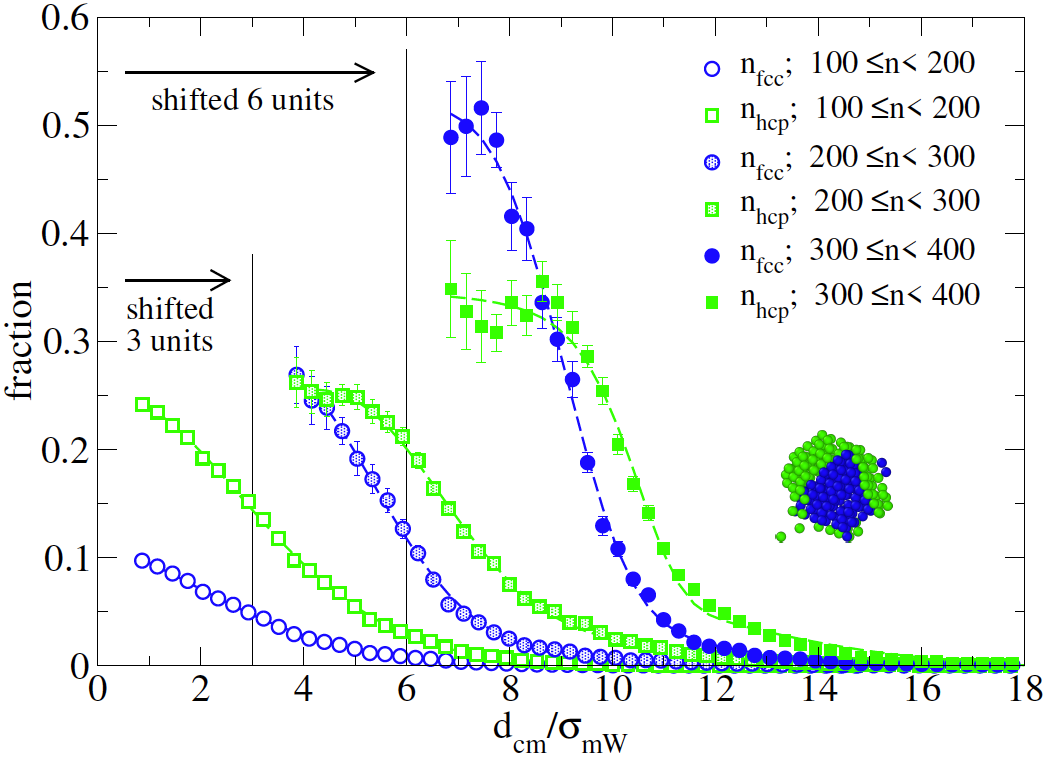}
                \caption{\label{fig:fraction_small} Average radial fractional composition of the main cluster of hard spheres for different range of cluster size as identified by LID. $d_{cm}$ is the distance from the center of mass of the cluster and $\sigma_{HS}$ the hard sphere diameter. Curves for $200\leq n <300$ and $300\leq n <400$ are shifted along the x-axis of 3 and 6 units respectively. Dashed fitting lines are a guide for the eyes. The snapshot shows the section of a typical nucleus of size $n\simeq 400$ particles. \emph{fcc} and \emph{hcp} particles are in blue and green respectively.}
        \end{center}
\end{figure}

\begin{figure}[!t] 
	\begin{center}
		\includegraphics[width=8.5cm]{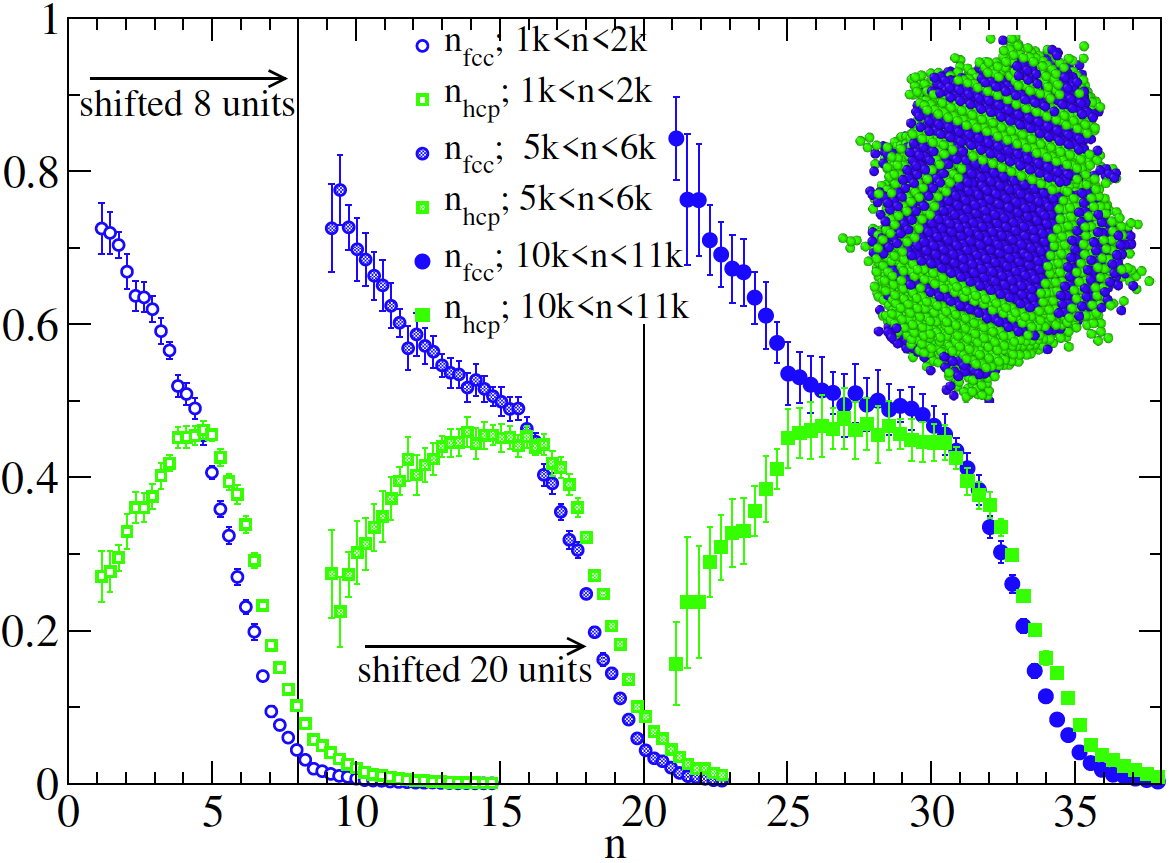}
                \caption{\label{fig:nc_HS_BIG} Average radial fractional composition of the main cluster of hard spheres for different range of cluster size as identified by LID. $d_{cm}$ is the distance from the center of mass of the cluster and $\sigma_{HS}$ the hard sphere diameter. Curves for $5000< n <6000$ and $10000< n <11000$ are shifted along the x-axis of 8 and 20 units respectively. The snapshot shows the section of a typical nucleus of size $n\simeq 11000$ particles. \emph{fcc} and \emph{hcp} particles are in blue and green respectively.}
        \end{center}
\end{figure}

The imbalance between the two polytypes, \emph{fcc} and \emph{hcp}, is measured with the ratio $r=n_{fcc}/n_{hcp}$, which we plot in the right-inset of Fig.\ref{fig:nc_HS}(f) as a function of the cluster size $n$. The ratio is not constant: it shows a predominance of \emph{hcp} for small values of $n$, which then converges towards an homogeneous composition as the size $n$ increases. As we noted in Fig.~\textit{1g}, at sizes above the critical value, nuclei are also not homogeneous, with the \emph{fcc} phase being more abundant on average towards the centre of the nucleus.

To understand the appearance of onion-like structures, in Fig.~\ref{fig:fraction_small} we plot the average radial fractional composition of crystalline clusters for different sizes, ranging from pre-critical nuclei to nuclei just above the critical size. The figure confirms that there is a transition between spatially uniform nuclei ($n\lesssim200$) where \emph{hcp} is the majority component, to larger nuclei where the core becomes more abundant in \emph{fcc} and the outer layers in \emph{hcp}.
Visual inspection of these nuclei reveals that the presence of an \emph{fcc}-rich core surrounded by stacking faults. 
There are two reasons for the size-dependent stability of \emph{fcc} cores. The first one is that \emph{fcc} is a cubic crystal, and thus can form stacking disorder along 4 independent directions (along the {1,1,1} planes) instead of only one direction as in the case of the \emph{hcp} crystal (which has hexagonal symmetry, and where the only stacking direction is the one perpendicular to the basal plane). The inset of Fig.~\ref{fig:fraction_small} shows a snapshot from the formation of these structures: an \emph{fcc} core (blue particles) developing stacking faults in two directions (green \emph{hcp} particles). The second reason is that the intersection of the stacking planes growing in different directions creates five-fold coherent grain boundaries from which the crystal can go radially maintaining an \emph{fcc}-rich core. These grain boundaries were first observed in Ref.~\cite{o2003crystal} for HS particles.

These observations are confirmed by looking at the radial fractional composition for large clusters, plotted in Fig.~\ref{fig:nc_HS_BIG}. With increasing size, the core of the nuclei retains the \emph{fcc}-rich character, while the surface develops an intermediate plateau with equimolar composition. This region is due to random stacking along one or multiple  {1,1,1} planes that emanate from the nucleus core. An example of this process is shown in the inset of Fig.~\ref{fig:nc_HS_BIG}.
The preference for \emph{hcp} in the outer-most part of the surface of nuclei shown by clusters of all sizes can be explained with the preference of clusters of tetrahedra in the liquid phase to coalesce via their faces in order to form locally dense aggregates \cite{anikeenko2007}, and this prevalent tetrahedral arrangement is compatible with the \emph{hcp} phase.

We now investigate the transition between pre-critical homogeneous nuclei and onion-like structures. To characterize the change in structure we compute the gyration tensor $S_{\alpha\beta}$ \begin{equation}
S_{\alpha\beta}=\dfrac{1}{2n^2}\sum_{i=1}^{n}\sum_{j=1}^{n}(r_{\alpha}^{i}-r_{\alpha}^{j})(r_{\beta}^{i}-r_{\beta}^{j})     
\end{equation}
where $\alpha,\beta=x,y,z$, and $r_{\alpha}^i$ is the $\alpha$ component of the position vector of the particle $i$ belonging to the cluster.
The eigenvalues of $S_{\alpha\beta}$ are also called principal moments and can be written as the ordered elements $\lambda_x^2\leq\lambda_y^2\leq\lambda_z^2$, and the radius of gyration is defined as $R_g=\sqrt{Tr(S)}=\sqrt{\lambda_x^2+\lambda_y^2+\lambda_z^2}$.

\begin{figure}[!t] 
	\begin{center}
		\includegraphics[width=5.3cm]{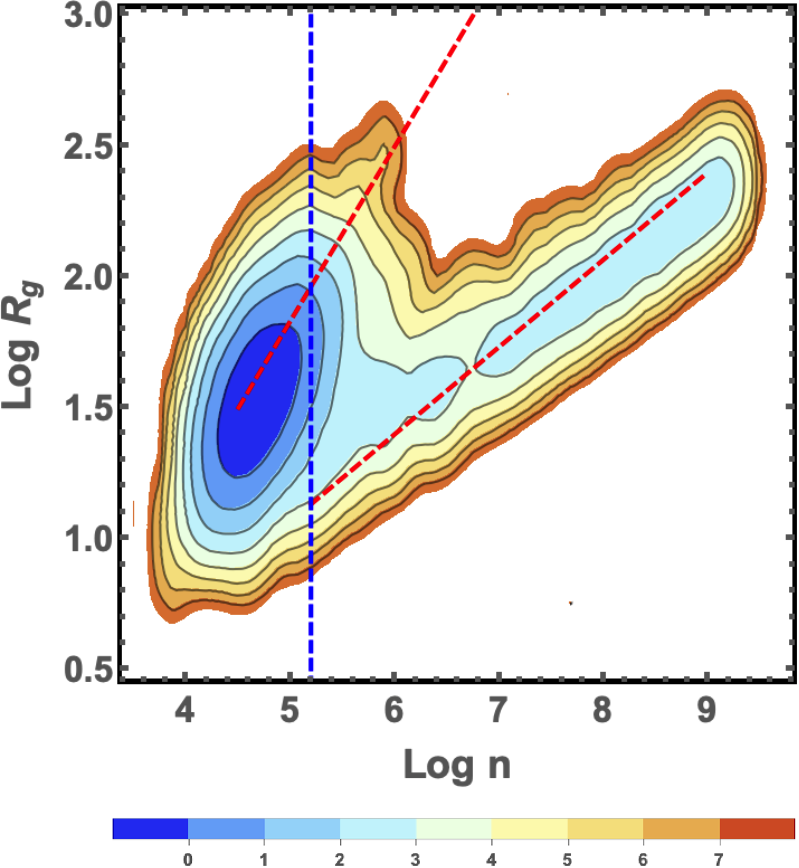}
		\includegraphics[width=3.1cm]{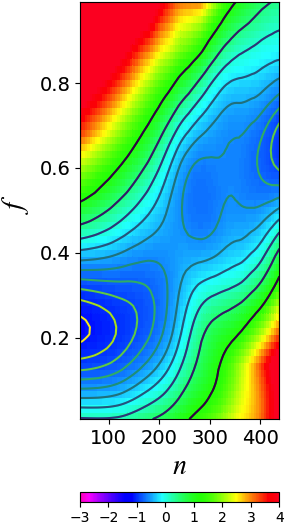}
		\caption{\label{fig:histograms} Histogram plots, $F(n,x)=-log P(n,x)$, where $P(n,x)$ is the reduced probability distribution function for the variables $n$ (size of the nucleus), and $x=R_g$ (radius of gyration) in the left panel and $x=f$ (\emph{fcc} composition ratio of the core) in the right panel.}
	\end{center}
\end{figure}

In Fig.~\ref{fig:histograms} we plot the normalized histograms 

$$
F(n,x)=-log P(n,x)
$$

where $P(n,x)$ is the reduced probability distribution function taken from our simulations data, with $n$ being the size of the nucleus, and $x=R_g$ (radius of gyration) in the left panel and $x=f=n^c_{fcc}/n^c$ (see definition in the following) in the right panel.

We first examine $x=R_g$ (left panel). Up to the critical nucleus size, $F(n,x)$ coincides with the potential of mean force for the two reaction coordinates $n$ and $x=R_g$.
The dashed blue line indicates the critical nucleus size, while the red dashed lines are power laws with $R_g\sim n^{2/3}$ (surface scaling) and $R_g\sim n^{1/3}$ (volume scaling) respectively. The figure shows that there is a clear distinction between pre-critical clusters, with large surface fluctuations, and post-critical clusters. Large surface fluctuations for small nuclei are compatible with previous experimental observations on repulsive colloids \cite{gasser2001,tan2014}. The majority of pre-critical nuclei are not compact enough for barrier crossing, and the path with the smallest barrier selects nuclei from the population with small radius of gyration (compact nuclei). This transition occurs in correspondence of the nucleus size where onion-like structures start to appear.
Indeed, stacking and defects like grain boundaries, which favour the formation of \emph{fcc} in the inner part of nuclei, can take place only when they are compact enough.

In the right panel of Fig.~\ref{fig:histograms} the reaction coordinate $x=f=n^c_{fcc}/n^c$ is given by the fraction of \emph{fcc} particles in the core of a nucleus, where the core is defined by a sphere of radius $3\sigma_{HS}$ centred in the barycentre. Results for different values of the sphere diameter are qualitatively similar. This choice in the computation of $f$ allows us to better highlight the transitions in the core for the small cluster sizes we are considering here. From the plot we can see a distinction between the \emph{fcc}-core-poor ($f<0.5$) basin at small $n$, and the \emph{fcc}-core-rich ($f>0.5$) basin at large $n$. Lines represent contour lines. The saddle point is found at a value of $n$ close to the estimated value of the critical nucleus.

Overall, Fig.~\ref{fig:histograms} shows that crystal nuclei that pass the nucleation barrier are more compact and have a higher \emph{fcc} content compared to pre-critical nuclei.

\subsection{$\mbox{m}$W water}

\subsubsection{Critical nucleus}
\label{sec:critical_nucleus}

\begin{figure}[!t] 
	\begin{center}
		\includegraphics[width=8.5cm]{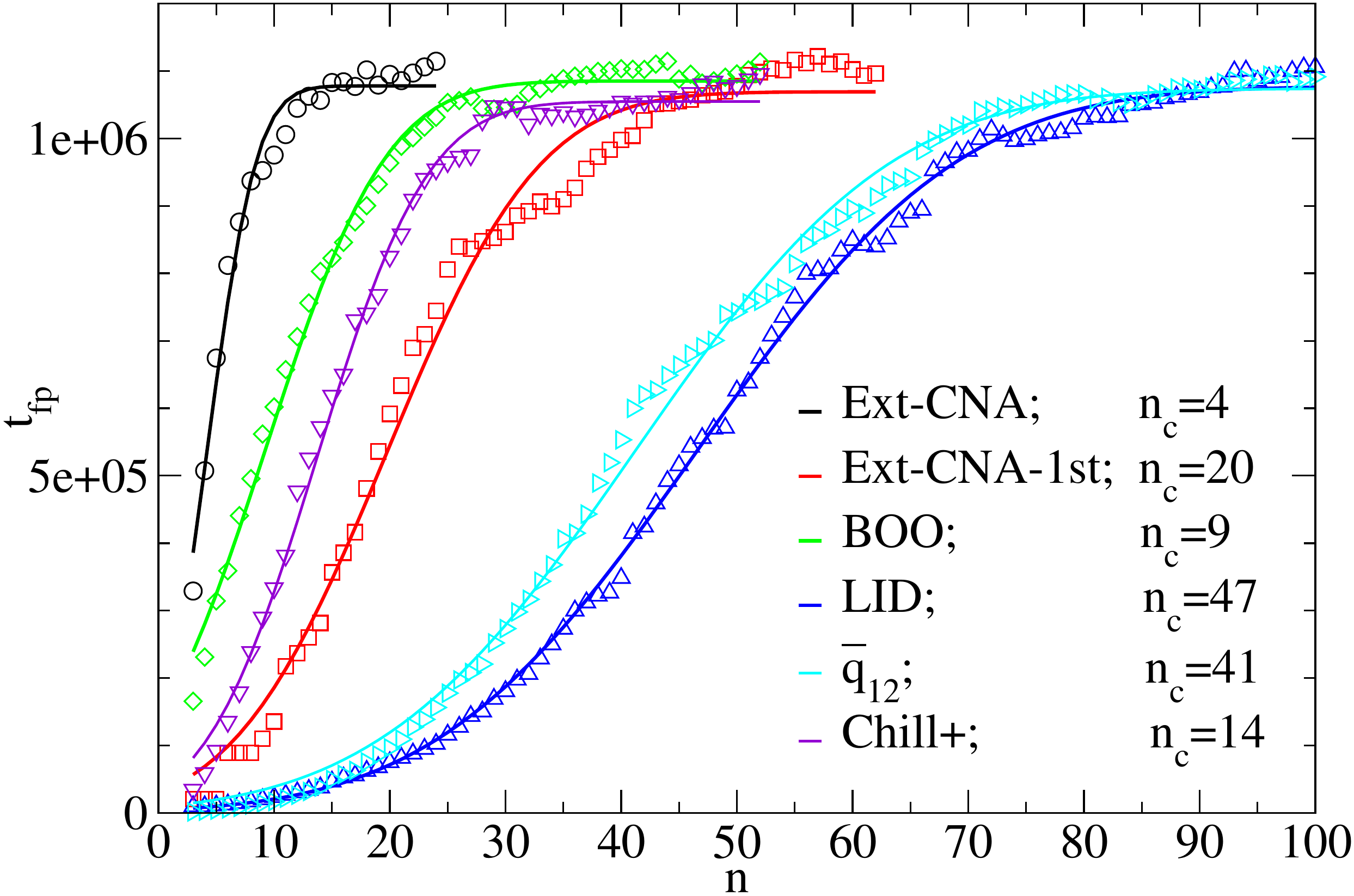}
		\caption{\label{fig:FPT} Average first passage time $t_{fp}$ as a function of nucleus size computed using different methods for local structure identification (see legend).}
	\end{center}
\end{figure}

First of all, we estimate the size of the critical nucleus by using the mean first passage theory \cite{wedekind2008,russo2013}.
This theory allows to estimate the average time at which the growing nucleus overcomes the nucleation barrier, and then to estimate the critical nucleus size $n_c$.
The mean first passage time $t_{fp}(n)$, which gives the average time after which a nucleus of size $n$ appears first in the system, is given by
\begin{equation}
t_{fp}(n)=\dfrac{1}{2kV}\{1+\mbox{erf}[Z(n-n_c)]\}    
\end{equation}
where $k$ is the nucleation rate, erf is the error function, and $Z=\sqrt{-\Delta G''(n_c)/(2\pi K_BT)}$ is the Zeldovich factor. $\Delta G''$ is the second derivative of the formation free energy of nuclei. $n_c$ corresponds to the value of $n$ where the curvature of $t_{fp}(n)$ changes its sign.
In Fig.\ref{fig:FPT} we show $t_{fp}(n)$ versus the nucleus size $n$. From it we can see a big variation in the estimation of $n_c$ from the different methods compared here.
To summarize these results, Ext-CNA, Ext-CNA-1st, Chill+ and BOO give a small value for $n_c$ going from 4 to 20. $\bar{q}_{12}$ and LID give values for $n_c$ close to each other, that is 41 and 47, respectively, while $\bar{q}_{4}\bar{q}_{6}$ gives a value for $n_c$ which is very sensitive to the protocol employed to compute it (see Appendix \ref{app:A}).

\subsubsection{Composition ratio of the nucleus}

\begin{figure}[!t] 
	\begin{center}
		\includegraphics[width=8.5cm]{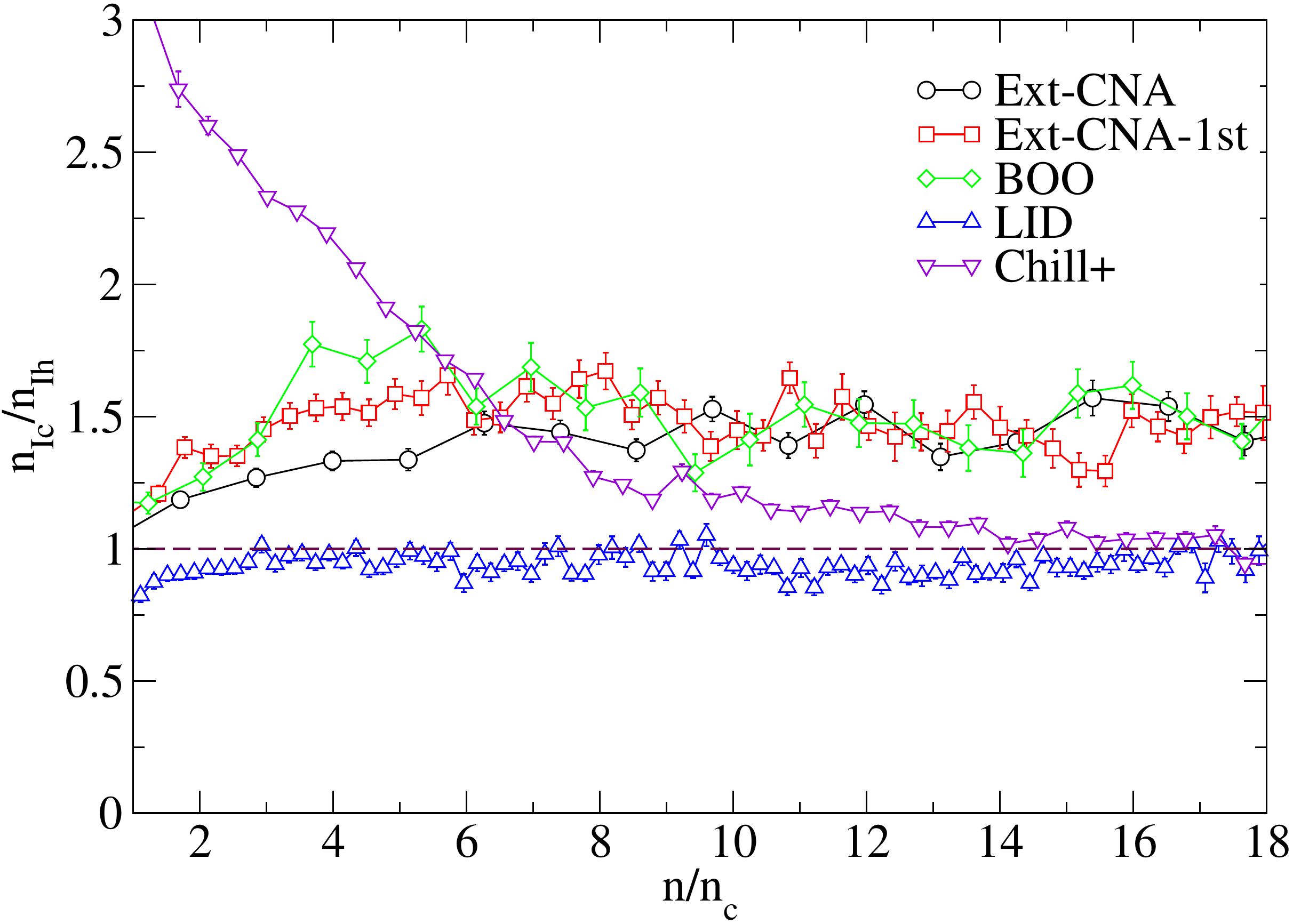}
		\caption{\label{fig:Comp} Average ratio $r=n_{I_c}/n_{I_h}$ between the number of particles composing the nucleus in the cubic phase (I$_c$) and the hexagonal phase (I$_h$) using different methods for local structure identification.
		$r$ is plotted against the nucleus size ($n$) normalized by the critical nucleus size of a specific method $i$ ($n_c^i$).}
	\end{center}
\end{figure}

After estimating $n_c$ we analyze the composition of the main cluster obtained from the different identification methods. 
In Fig.~\ref{fig:Comp} we show the ratio $r=n_{Ic}/n_{Ih}$ between the number of particles belonging to the main cluster which are associated to the cubic ice ($n_{Ic}$) and those associated to the hexagonal ice ($n_{Ih}$) versus the normalized nucleus size $n/n_c$, where $n_c$ is the critical size of the nucleus given by the method under consideration. 
We do not show the ratio of particles of the nucleus identified as ice~0 because only some of the methods analyzed here include it between the possible crystal phases.

%

As shown by Prestipino in Ref.~\cite{prestipino2018}, the $\bar{q}_4\bar{q}_6$ method can give predictions on the composition of the nucleus completely different by changing the protocol used to compute and partition this 2-dimensional OP.
Here we consider different protocols obtaining different values for the ratio $r$ and report the results separately in Appendix \ref{app:A}.

Ext-CNA and Ext-CNA-1st give a preference to I$_c$ with a value of the ratio $r$ between $1.3$ and $1.4$.  
BOO predicts a value $r\sim 1.4$ for small normalized nucleus size, while for larger normalized nucleus size it drops to values closer to 1 ($r\sim 1.1$). Chill+ has a strong imbalance towards ice I$_c$ for small sizes and reaches $r\sim 1$  only for large nucleus size.
Only LID measures $r\sim 1$, except for small cluster size, where the hexagonal ice becomes predominant, a similar behavior to what we observed for hard spheres (see Sec.\ref{sec:small_HS}). 
As shown in Sec.\ref{sec:benchmark}, the ratio $r\sim 1$ given by LID and Chill+ at large nucleus size is not observed in other methods. The larger value of $r$ of these methods comes from the fact that they perform well only near the centre of the nucleus, which comprises a majority of cubic ice environments, and perform worse near the surface, where the hexagonal environments are more abundant than cubic ones.

\subsubsection{Radial composition of the nucleus}
\label{sec:radial_comp}

\begin{figure}[!t]
\centering
\includegraphics[angle=0, width=8.5cm]{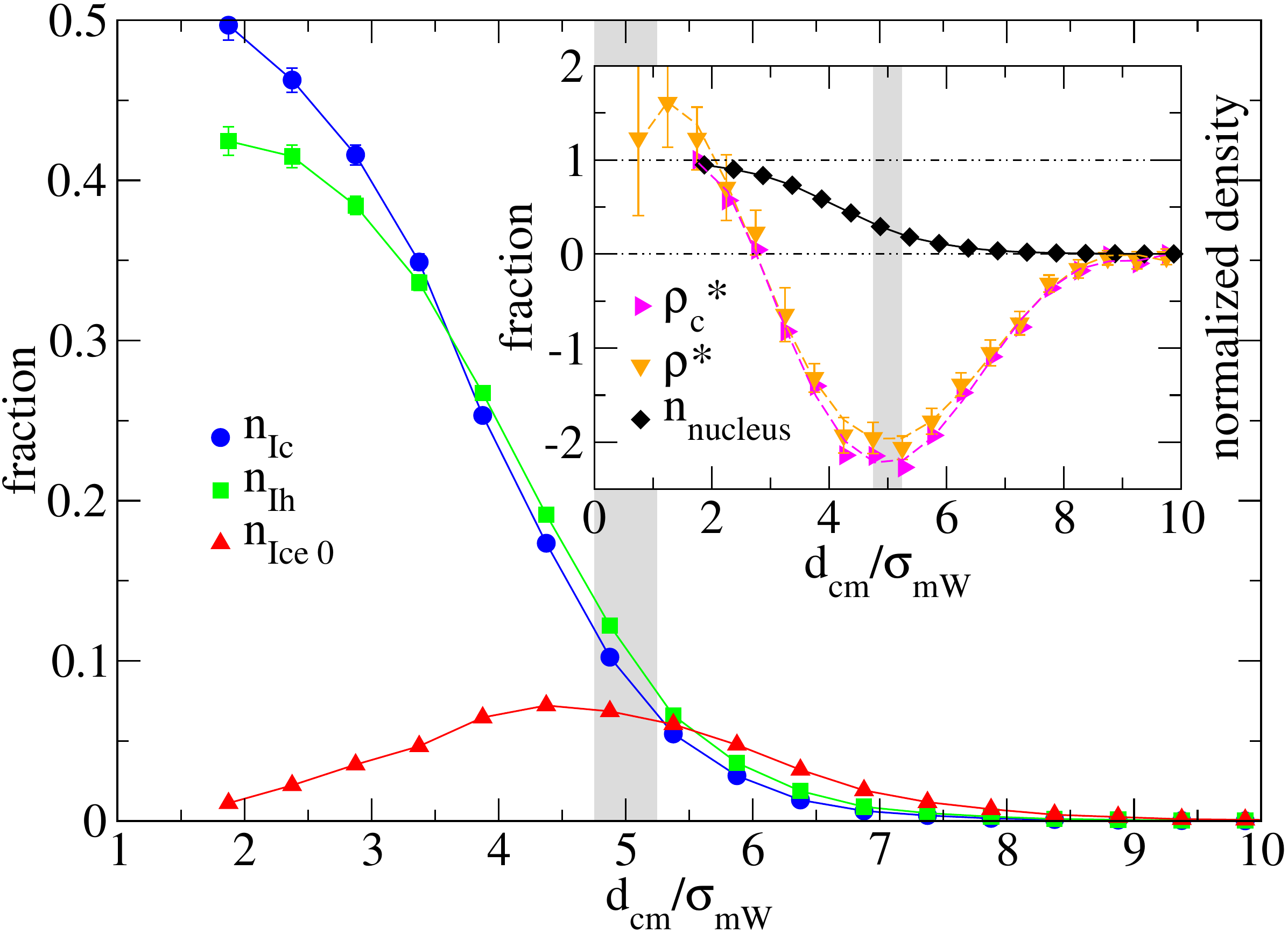}
    \caption{\label{fig:nc_Water} Homogeneous nucleation of mW water. Average radial fraction composition of the main cluster (for clusters of size $150\leq n\leq 200$) as identified by LID. $d_{cm}$ is the distance from the center of mass of the cluster and $\sigma_{mW}$ the mW water molecule diameter. The inset shows the normalized density profiles (colored symbols) for the same nuclei considered in the main panel, while black symbols represent the total fraction of all crystalline particles.}
\end{figure}

Fig.~\ref{fig:nc_Water} shows the average radial composition for nuclei of size $150\leq n\leq 200$ obtained using LID. We find the same nucleation property that also characterized HS nuclei: while the overall average composition is the same between the stable ice I$_c$ and I$_h$ polytypes ($r\sim 1$), the cubic diamond is more abundant than the hexagonal diamond near the centre of the nucleus. But the mW model also offers additional insights respect to the HS system.
LID is the only method to detect the presence of ice~0 like structures (red symbols), whose growth is slower than the volume growth of both the ice $I$ polytypes (See Fig.~\ref{fig:nc_Water_A}). Fig.~\ref{fig:nc_Water} indeed confirms the presence of a small population of $0$-like environments which peaks towards the surface of the nucleus. An independent confirmation of this unusual surface behaviour of mW water can be seen in the inset of Fig.~\ref{fig:nc_Water} where we plot (orange symbols) the normalized density $\rho^*=(\rho-\rho_{f})/(\rho_{x}-\rho_{f})$, where $\rho_{x}=0.985$~g/cm$^{3}$ is the bulk density of the ice $I$ phase and $\rho_{f}=0.980$~g/cm$^{3}$ is the density of the bulk liquid phase, at the thermodynamic conditions considered here. Importantly, the density of ice~0 ($\rho=0.953$~g/cm$^{3}$) is lower than both the metastable liquid and ice~$I$ crystals at the same conditions. Indeed we observe that, instead of monotonically increasing from $\rho_{f}$ at the surface towards $\rho_x$ at the center of the nucleus, the density profile has a very pronounced density minimum towards the surface of the nucleus. The location of this minimum (which is computed independently from any structural order parameter, if not for the location of the center of mass) corresponds exactly to the location of the maximum in the ice~0 population (a grey vertical band is drawn in Fig.~\ref{fig:nc_Water} to highlight the location of both). To further support the association between the density minimum and the presence of a population of low-density local structures, we have independently computed the local density of particles associated with each environment, and in the inset of Fig.~\ref{fig:nc_Water} we plot the density $\rho_c^*$ obtained by weighting these local densities with the fractional compositions obtained from LID (main panel). We see that $\rho_c^*$ exactly mirrors $\rho^*$, showing that we have obtained a good partial density decomposition. These results offer an even stronger case for the onion-like structure of growing nuclei, which in the case of water appears to be multi-step.

\begin{figure}[!t]
\centering
\includegraphics[angle=0, width=8.5cm]{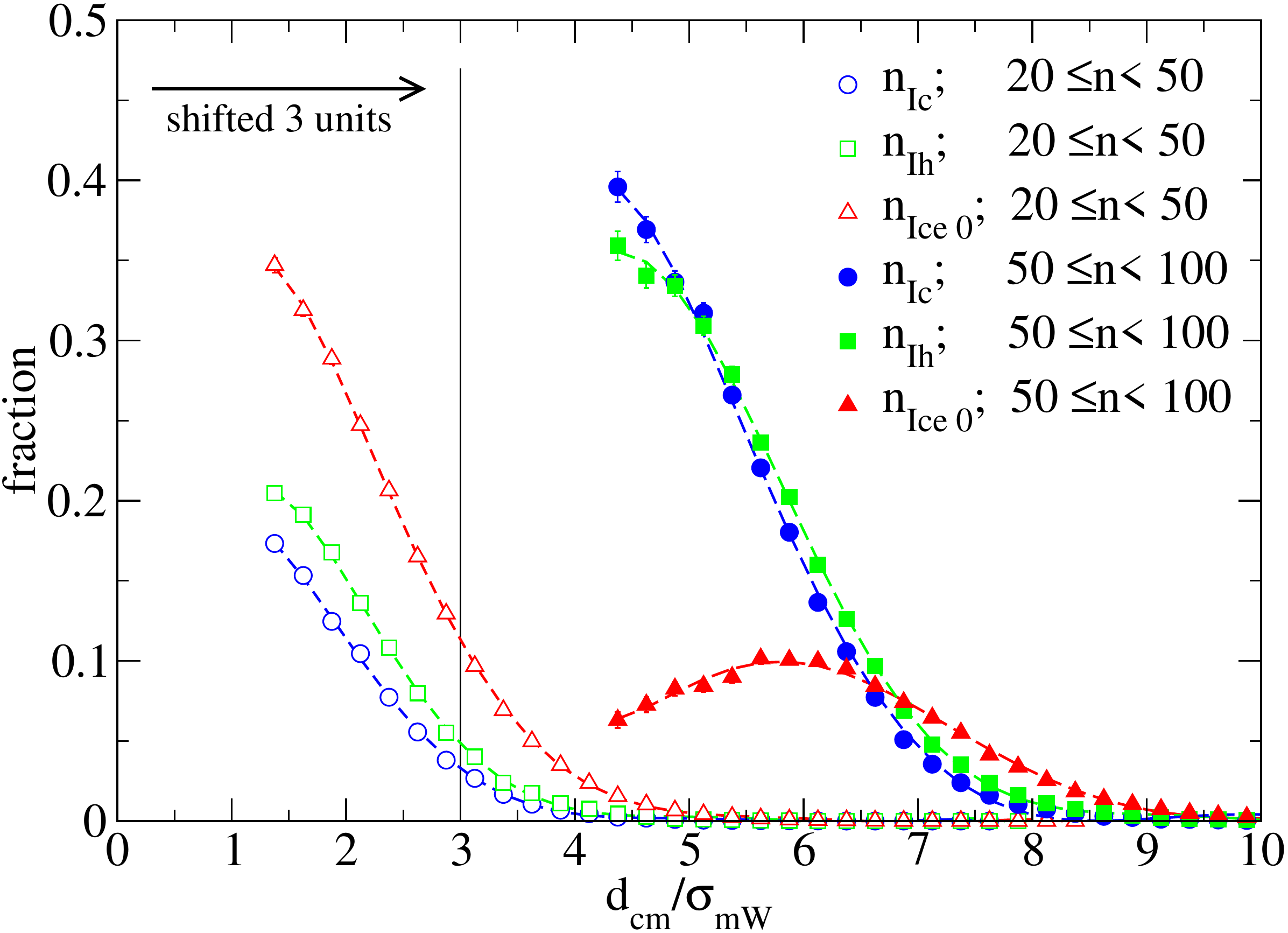}
    \caption{\label{fig:fraction_Water} Average radial fraction composition of the main cluster for clusters of size in the range 20-50 and 50-100 (shifted along the x-axis of 3 units), as identified by LID. $d_{cm}$ is the distance from the center of mass of the cluster and $\sigma_{mW}$ the mW water molecule diameter.}
\end{figure}

The presence of onion-like structures, and their radial composition is not explained by the small free energy differences between the bulk phases.
In fact we observe that the cubic crystals (\emph{fcc} and I$_c$) are found more abundantly near the centre of the nucleus, while their hexagonal counterparts (\emph{hcp} and I$_h$) are found more abundantly towards the surface. In terms of bulk free energies instead the stable phases are \emph{fcc} and ice  I$_h$ in hard spheres and mW water respectively. To account for the ordering of the phases one needs to consider the free energy cost of structural fluctuations, which is size-dependent. It has been observed that small finite-size clusters of the cubic phase gain relative stability compared to the hexagonal phase thanks to the entropy associated with stacking disorder~\cite{molinero2017nature,o2003crystal,li2011homogeneous} and the low energetic cost of their grain boundaries~\cite{leoni2019}.

We repeat here the analysis of small (pre-critical) clusters that we performed for HS (see Sec.\ref{sec:small_HS}).
In Fig.~\ref{fig:fraction_Water} we show the average radial fraction composition of the main cluster for two size range. For clusters of size in the range $20\leq n<50$ the nucleus is composed of a mixture of I$_c$, I$_h$ and ice~0 with predominance of ice~0 and then of I$_h$.
Going from pre-critical to immediately critical clusters, that is for clusters of size in the range $50\leq n<100$, the onion-like structure starts to appear with ice~0 forming a peak which shifts towards outer layers for increasing cluster size.
Also for mW water, as seen for HS, there is a selection of more compact clusters at the onset of the post-critical regime.

\subsubsection{Equilibrium trajectories}
To exclude the possibility that the nucleation pathway is due to the non-equilibrium nature of nucleation events at high supercooling, we apply the same analysis to trajectories obtained from Umbrella Sampling (US) simulations.
Umbrella sampling, and other techniques such as metadynamics or forward flux sampling, are usually employed in homogeneous nucleation to enhance the sampling of crystalline cluster \cite{quigley2008,li2011homogeneous,reinhardt2012,reinhardt2013,leoni2019}.
In order to test the LID OP against homogeneous nucleation in mW water, which would confirm its ability to capture the local crystalline phases I$_c$, I$_h$ and ice~0, we bias the umbrella sampling simulations by using LID as reaction coordinate.
For performance reasons, here we construct LID by considering spatial first and second neighbors of a local particle, as done for hard spheres, instead of energetic neighbors.
US simulations are performed with $N=10000$ mW particles at ambient pressure and $T=218$~K. 
\begin{figure}[!t]
\centering
\includegraphics[angle=0, width=8.5cm]{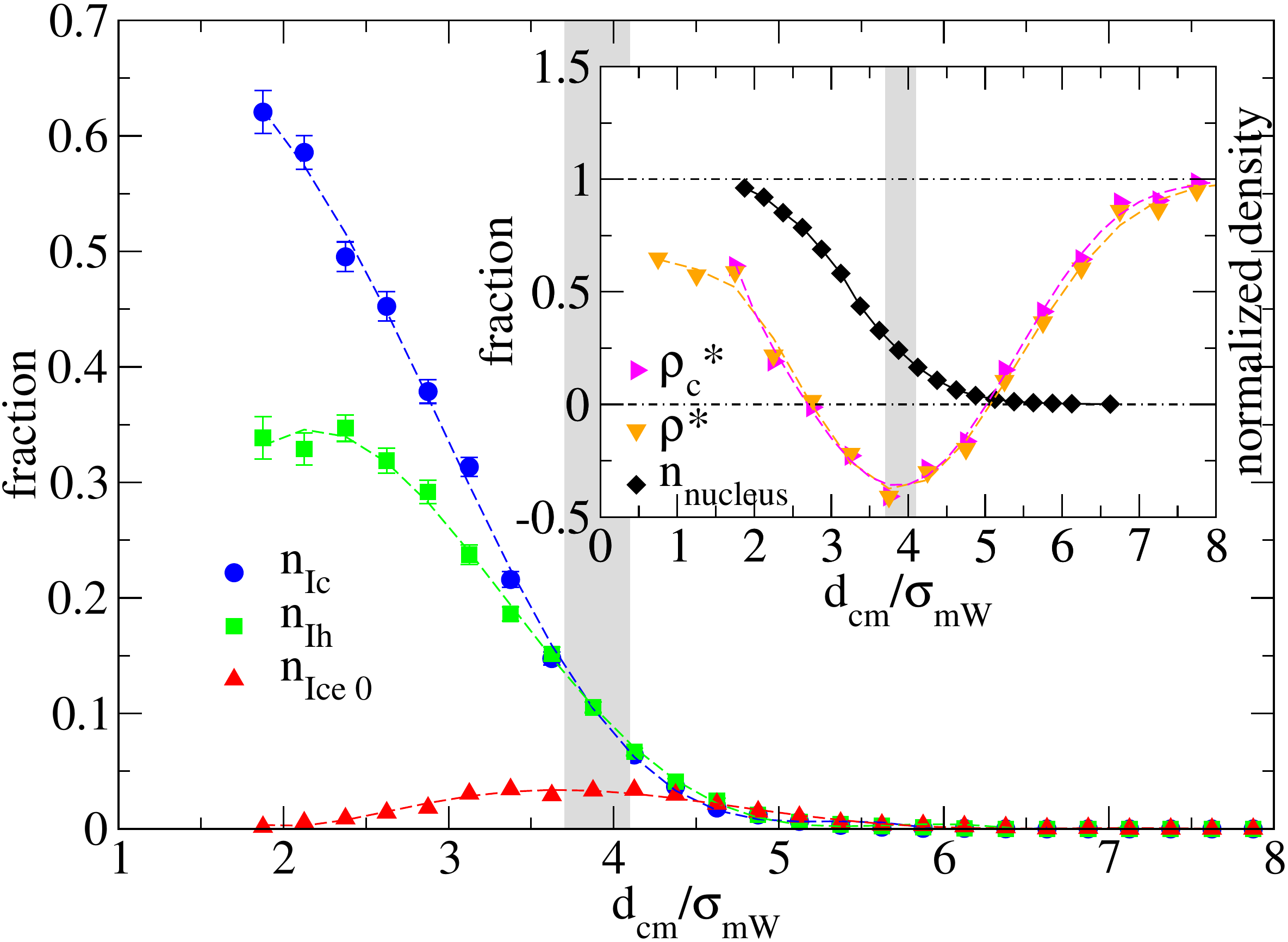}
    \caption{\label{fig:nc_Water_US} Homogeneous nucleation of mW water from LID-biased umbrella sampling. Average radial fraction composition of the main cluster (for clusters of size $50\leq n\leq 100$) as identified by LID with neighbors identification from spatial condition (see text). $d_{cm}$ is the distance from the center of mass of the cluster and $\sigma_{mW}$ the mW water molecule diameter. The inset shows the normalized density profiles (colored symbols) for the same nuclei considered in the main panel, while black symbols represent the total fraction of all crystalline particles. Dashed fitting lines are a guide for the eyes.}
\end{figure}

In Fig.~\ref{fig:nc_Water_US} we show the average composition of the main cluster nucleated with US simulations for clusters of size $50\leq n\leq 100$, as identified by the LID OP with spatial neighbors.
Differently from the spontaneous nucleation pathways analyzed before, US simulations allow us to study the structure of the nuclei in equilibrium. Moreover it allows us to study nucleation at higher temperatures (where spontaneous nucleation would not be observed). Despite these differences, we get a very similar result to that obtained by using LID from spontaneous nucleation (see Fig.~\ref{fig:nc_Water}): I$_c$ particles are more concentrated near the center of mass of nuclei, whereas I$_h$ particles are slightly more abundant near the surface (note that small differences in the fraction composition between phases are magnified when computing the number of particles in a crystalline phase composing the nucleus because it depends on the square of their distance from the center of mass), and ice~0 particles concentrated around the surface of nuclei. In the Inset of Fig.~\ref{fig:nc_Water_US} we show, as in Fig.~\ref{fig:nc_Water} for spontaneous nucleation simulations, the total fraction of crystalline particles (black diamonds), the normalized density $\rho^*=(\rho-\rho_x)/(\rho_f-\rho_x)$ (orange pointing downwards triangles), where $\rho_f=0.995$~g/cm$^3$, $\rho_x=0.983$~g/cm$^3$ and $\rho_{ice~0}=0.952$~g/cm$^3$ at the present thermodynamic conditions, and the normalized density $\rho_c^*$ (magenta pointing rightwards triangles) computed by weighting the local densities of each phase with their fractional compositions obtained from LID (main panel).
Note that here the linear transformation applied to $\rho$ in order to get a normalized density $\rho^*$ differs from the one used in Fig.~\ref{fig:nc_Water} for the swap of $\rho_f$ with $\rho_x$ because at $T=218$~K $\rho_f>\rho_x$, while at $T=204$~K it is the opposite (see Ref.~\cite{leoni2019}).

\begin{figure}[!t]
\centering
\includegraphics[angle=0, width=8.5cm]{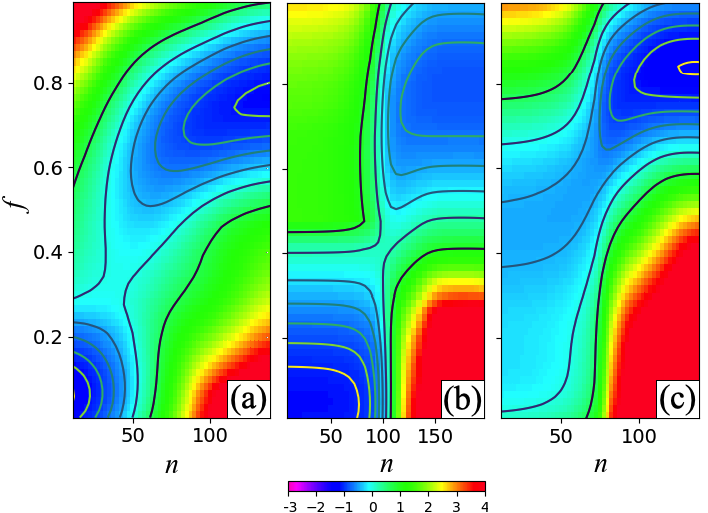}
    \caption{\label{fig:FES_Water_US} 2d-plots of $F(n,f)=-log[P(n,f)]$ for direct molecular simulations (a), and US simulations with configurations initialized using the seeding technique to have a nucleus in the I$_{sd}$ (b) or I$_c$ (c) phase.}
\end{figure}

Similarly to the HS case (right panel of Fig.~\ref{fig:histograms}), in Fig.~\ref{fig:FES_Water_US} we show 
the normalized histograms of $F(n,f)=-log[P(n,f)]$. The reaction coordinate $f$ is defined in the same way as for HS where the radius of the sphere defining the core is now $3\sigma_{mW}$.
Panel $(a)$ shows direct molecular simulations, while panels $(b)$ and $(c)$ are the result of US simulations, again at $T=218K$, in which we initialized the configurations using the seeding technique \cite{bianco2021} with nuclei in the I$_c$ and I$_{sd}$ phase respectively. I$_{sd}$ is the stacking disordered phase. For details on the simulation procedure see Ref.~\cite{leoni2019}.
From panel $(a)$ we can see the presence of the two basins, the I$_c$-core-poor ($f<0.5$) at small $n$, and the I$_c$-core-rich ($f>0.5$) at large $n$, separated by the saddle point located in correspondence of the critical nucleus (at $T=218 K$ $n_c$ is close to $\sim 100$).
The US simulations (panels (b) and (c)) offer a view on the equilibrium landscape of the nucleation process for the formation of different nuclei: I$_{sd}$ nuclei in panel (b), and I$_c$ nuclei in panel (c). The potential of mean force for the I$_{sd}$ nucleation shows two channels: one corresponding to a I$_c$-core-poor nuclei at small n, and one with I$_c$-core-rich nuclei at large n. The overall process in this case is similar to the one observed in direct simulations (panel (a)). The potential of mean force for the I$_c$ nuclei in panel (c) displays a process devoid of the I$_c$-core-poor basin, showing the existence of well-separated nucleation channels~\cite{james2019phase}.

\subsubsection{Dynamical behavior} 

\begin{figure}[!b] 
	\begin{center}
		\includegraphics[width=8.5cm]{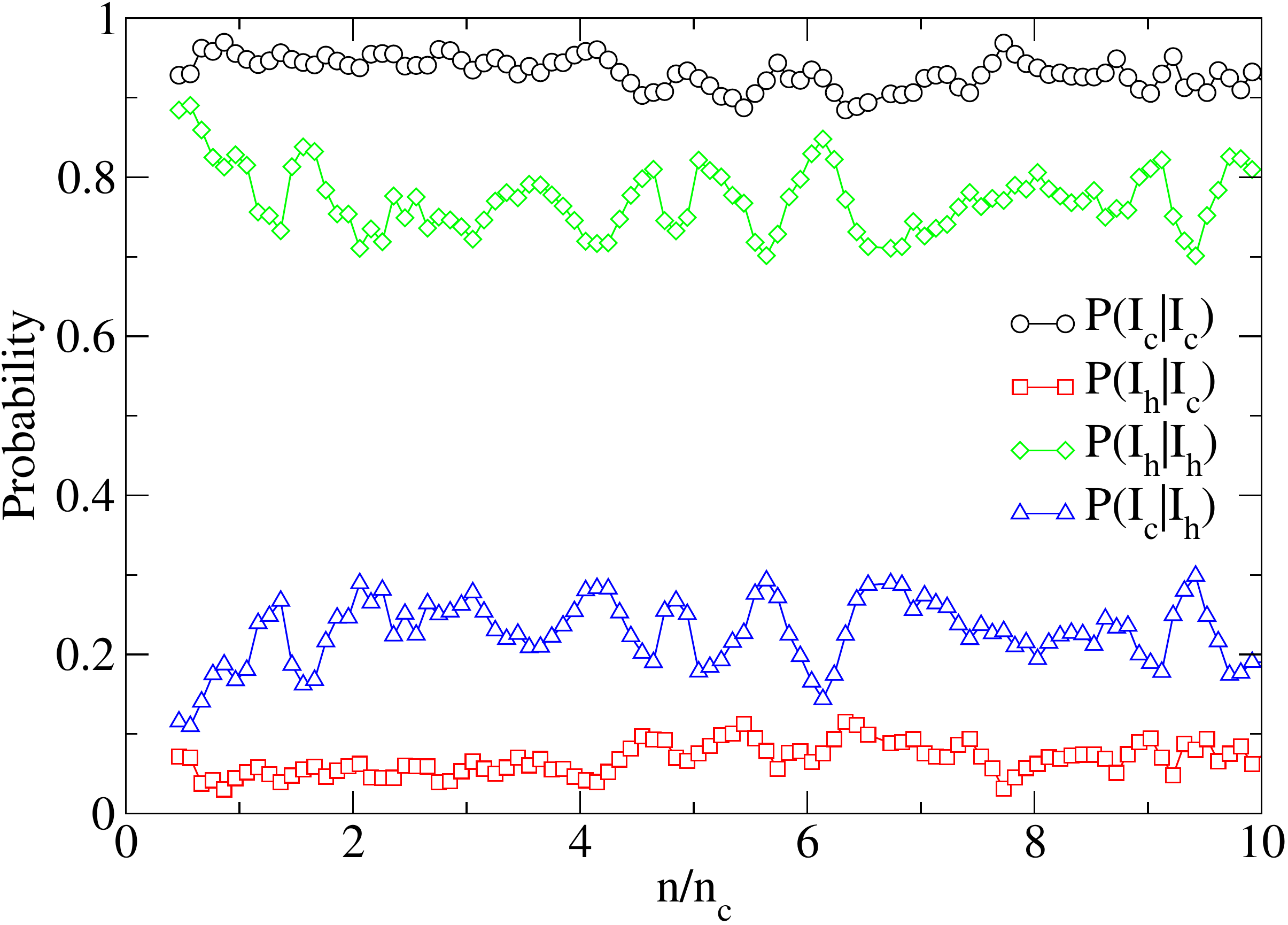}
		\caption{\label{fig:Trans} Conditional probability that a particle attaching to a nucleus of size $n$ will - stay in the same phase until the end of the dynamics (black circles for I$_c$, and green diamonds for I$_c$) - change phase at the end of the dynamics (red squares for I$_c$ which transforms into I$_h$, and blue triangles for I$_h$ which transforms into I$_c$).}
	\end{center}
\end{figure}

To study the dynamical behavior of the growing nucleus we compute how many particles attaching to the nucleus change their phase and how many do not during the entire dynamical process as a function of the nucleus size.
In particular, we trace the evolution of particles in the main cluster in reverse time: for each trajectory we count how many particles of the main cluster which are in a specific phase at the end of the dynamics are still found to be in that phase at the time when they attached to the cluster as a function of the cluster size $n$ at that time.
In Fig.~\ref{fig:Trans} we show the conditional probability that a particle in a cluster of size $n$ will stay always in the I$_c$ (I$_h$) phase during the whole dynamics, indicated with blue circles (green squares), and the conditional probability that at the end of the dynamics it will be in the opposite phase, indicated with black diamonds (red triangles). In this case we use the LID method to identify the local structure around each particle.
%

\begin{figure}[!t] 
	\begin{center}
		\includegraphics[width=8.5cm]{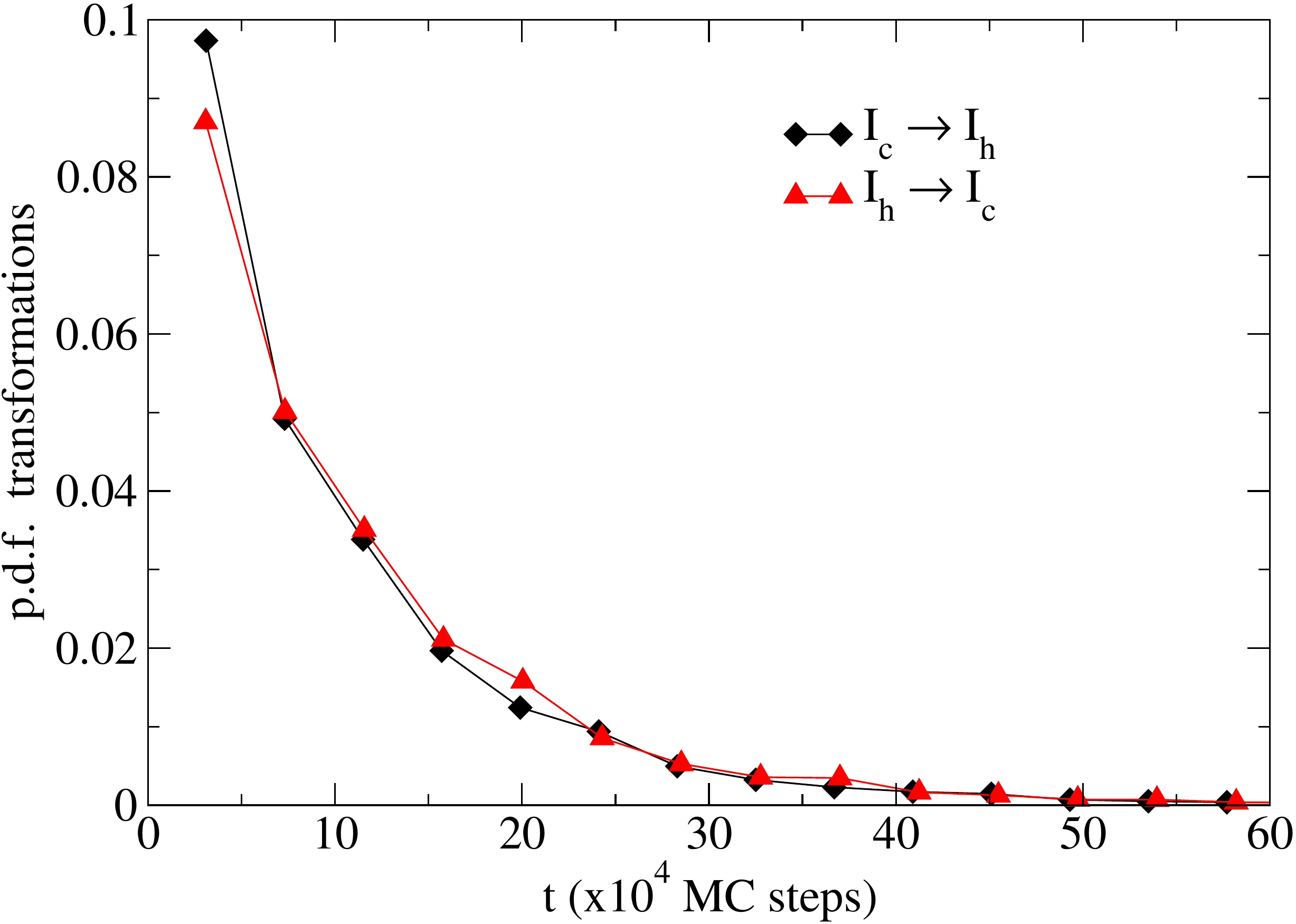}
		\caption{\label{fig:pdf_t} Probability distribution function (p.d.f.) of phase transformation versus time $t$ in $10^4$ MC steps units.}
	\end{center}
\end{figure}

From Fig.~\ref{fig:Trans} we can see that for critical clusters (that is for $n/n_c>1$) on average a particle appearing in the main cluster of size $n$ with phase I$_c$ (I$_h$) will stay in that phase for the whole dynamics with a conditional probability $p(I_c|I_c)\simeq0.93$ ($p(I_h|I_h)\simeq0.77$). Also the probability of starting with a phase and ending with the other phase is not symmetric: particles appearing in the main cluster of size $n$ with phase I$_c$ (I$_h$) will end up to be in the I$_h$ (I$_c$) phase with a conditional probability $p(I_h|I_c)\simeq0.07$ ($p(I_c|I_h)\simeq0.23$). 

We have seen that hexagonal local environments (more abundant on the surface) are more likely to change to cubic local environments as they get incorporated in the nucleus during the growth stage. To confirm that this transformation occurs on the surface, i.e. soon after local environments become crystalline, in Fig.~\ref{fig:pdf_t} we compute the probability distribution of the time between the first appearance of the crystalline environment (black diamond symbols for I$_c$ and red triangles for I$_h$) and its last phase transformation. We see that transformations occur exponentially fast in time following the same curve for both transformations, and are thus surface events.

\subsubsection{Precursors}

\begin{figure}[!t] 
	\begin{center}
		\includegraphics[width=8.5cm]{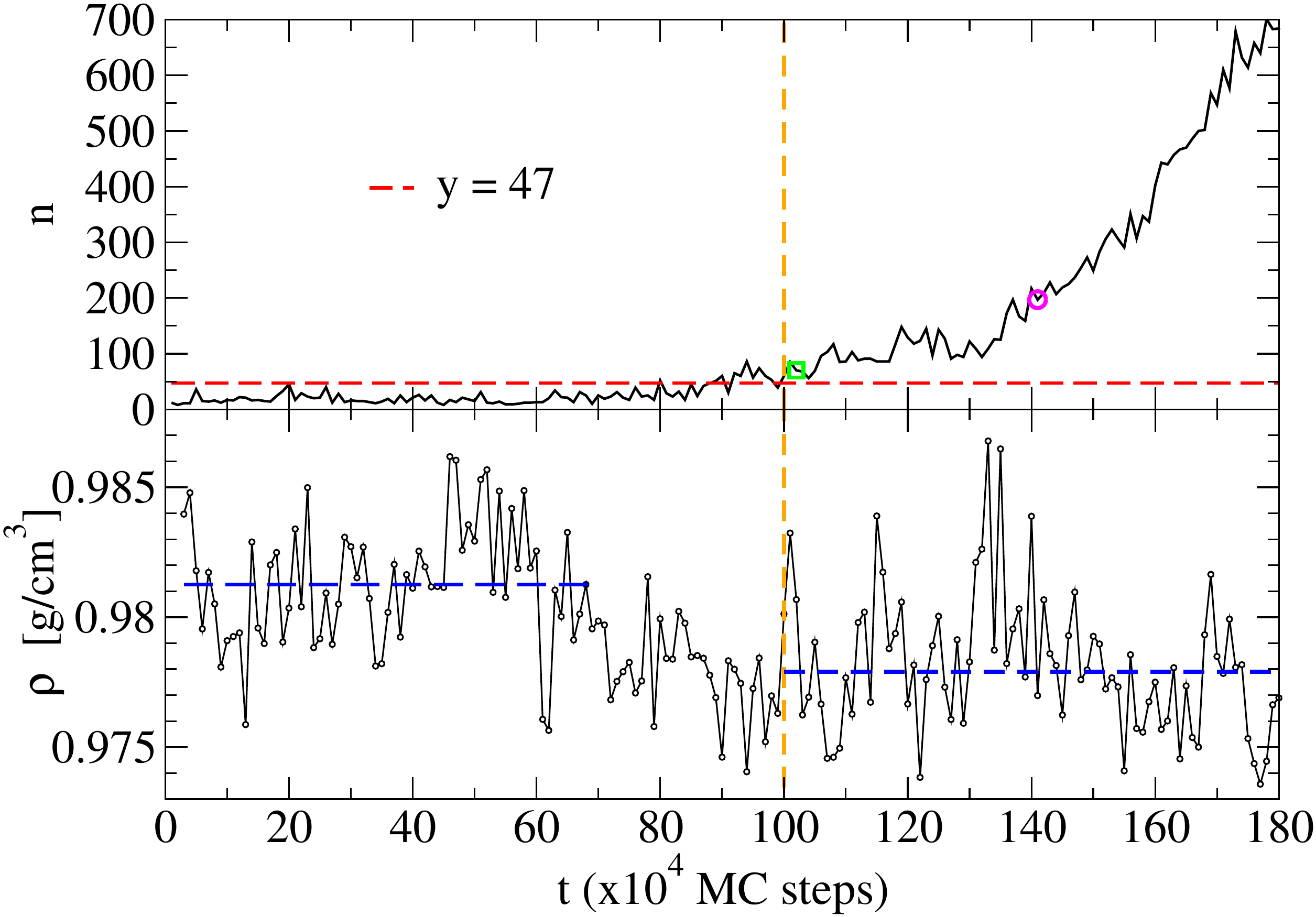}
		\caption{\label{fig:N_vs_t} Upper panel: size of the main cluster $n$, as identified by LID, versus time $t$ in $10^4$ MC steps units for a specific nucleation trajectory. The horizontal dashed red line corresponds to the critical nucleus size $n_c=47$, as identified by LID. The vertical dashed orange line corresponds to the largest time at which $n=n_c$. The green square and violet circle corresponds to the points (102,70) and (141,197), respectively.
		Lower panel: system density $\rho$ versus time $t$ in $10^4$ MC steps units for the same nucleation trajectory considered in the upper panel. The vertical dashed orange line corresponds to the largest time at which $n=n_c$. The horizontal dashed blue lines are obtained from density averages over a short time interval, and are a guide for the eyes.}
	\end{center}
\end{figure}
%
\begin{figure}[!ht] 
	\begin{center}
		\includegraphics[width=8.5cm]{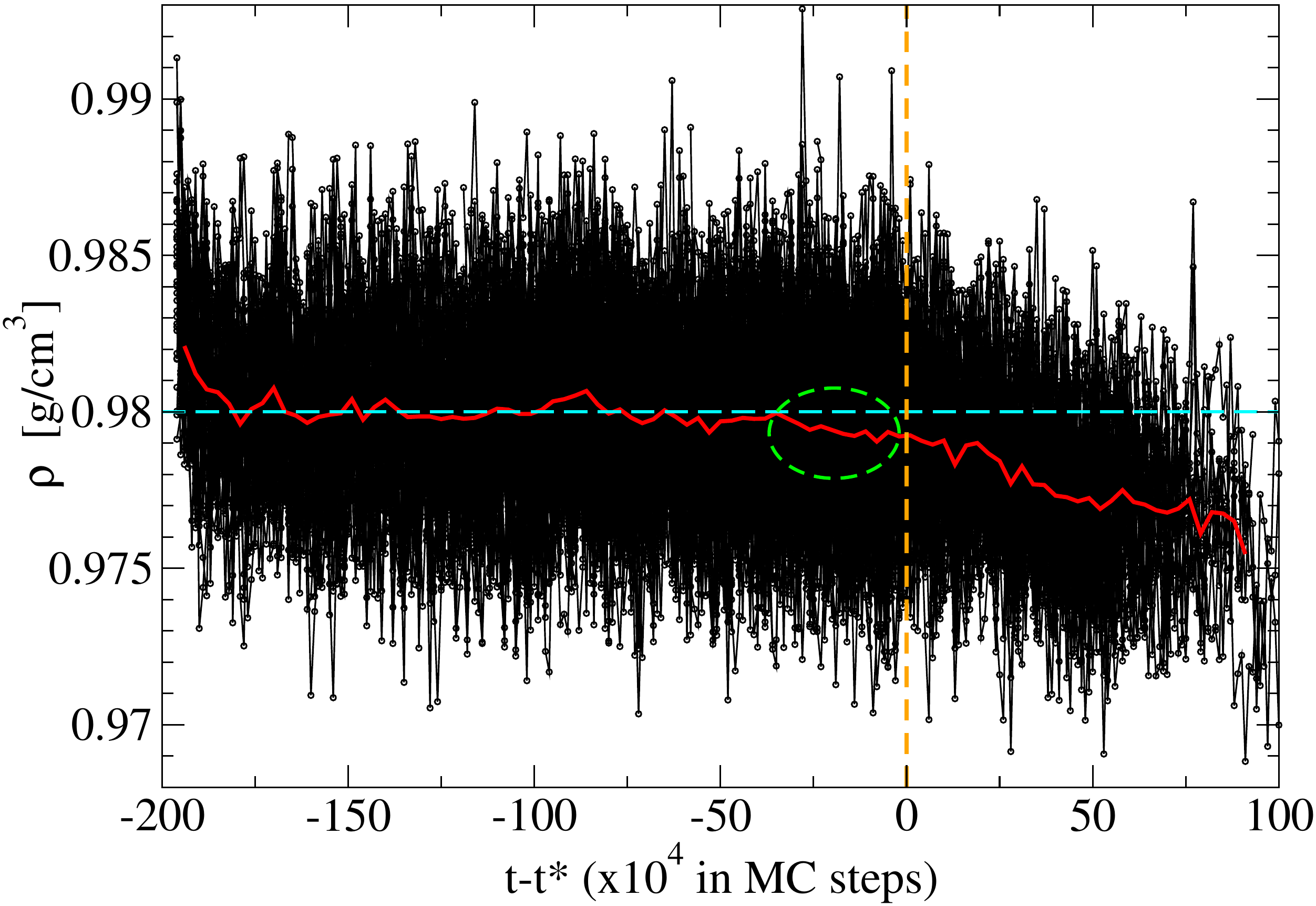}
		\caption{\label{fig:RHO_AV} System density $\rho$ of all nucleation trajectories as a function of $t-t^*$ in $10^4$ MC steps units, where $t^*$ is the last time (vertical dashed orange line) at which the main cluster in the system (identified by using LID) has critical size ($n=n_c$). $t^*$ is different for every trajectory $i$. The horizontal dashed cyan line represents the bulk liquid density ($\rho_L=0.980$g/cm$^3$ at the present thermodynamic conditions). The red line is the average of all densities. The green dashed circle highlight a pre-critical precursor region.}
	\end{center}
\end{figure}

Here we investigate the nature of the density decrease in proximity of the surface of the nucleus as found in the radial compositions of Figs.~\ref{fig:nc_Water},~\ref{fig:nc_Water_US}.
In the upper panel of Fig.~\ref{fig:N_vs_t} we show the size $n$ of clusters identified by LID as a function of MC steps for a specific trajectory. The horizontal dashed red line corresponds to the critical nucleus size $n_c=47$. We define $t^*$ as the time when the nucleus has the critical size $n=n_c$ for the last time during the growth process (vertical dashed orange line in the figure).
In the lower panel of Fig.~\ref{fig:N_vs_t} we show the system density $\rho$ as a function of MC steps for the same trajectory considered in the upper panel. The horizontal dashed blue lines are obtained from density averages over a short time interval, and highlight that $\rho$ decreases in correspondence of the formation of the critical nucleus.
At the thermodynamic conditions we consider here ($P=0$~Pa, $T=204$~K) the density of ice I, ice~0 and liquid phase is $\rho=0.984, 0.953, 0.980$~g/cm$^3$, respectively (see Ref.~\cite{leoni2019}). 
Differently from classical predictions, for which the formation of a crystalline nucleus should correspond to an increase in density at the present thermodynamic conditions, here we see the opposite. As discussed in Sec.\ref{sec:radial_comp} this density decrease can be explained by the formation of ice~0-like local structures in correspondence of the nucleus surface (see Fig.~\ref{fig:nc_Water}).
  
The same trend is observed in all nucleating trajectories $i$: in Fig.~\ref{fig:RHO_AV} we plot the densities as a function of the time from $t^*_i$.
In Fig.~\ref{fig:RHO_AV} the red line is the average density $\langle\rho(t-t^*_i)\rangle$, the horizontal dashed cyan line shows the bulk liquid density $\rho_L=0.980$~g/cm$^3$ at $T=204$~K, the vertical dashed orange line corresponds to $t=t^*$ for each trajectory $i$, and the dashed green circle indicates pre-critical nuclei. 
From Fig.~\ref{fig:RHO_AV} we see that the average density steadily decreases from pre-critical precursor regions.

\section{Conclusions}
\label{sec:Conclusions}

In two-step nucleation an intermediate phase is in a size-dependent competition with the stable phase (\emph{fcc} vs \emph{hcp} in hard spheres or cubic ice vs hexagonal ice in mW water)\cite{debenedetti1996metastable,kelton2010nucleation,james2019phase}.
Here we consider this phenomenology in the particular case of polytype nucleation. We have studied the microscopic nucleation pathway in systems characterized by a competition of different polytypes, whose bulk free energy properties don't discriminate between them. Even in systems where no classical argument for a two-step process is expected, we find a 
selection of critical clusters with a compact structure that leads to the formation of onion-like structures, thus considerably extending the number of system showing this type of nucleation mechanism~\cite{toth2011amorphous,santra2013nucleation,barros2013liquid,russo2016nonclassical,tang2017competitive,lutsko2019crystals}. In particular, our results highlight the role of structural fluctuations in nucleation phenomena~\cite{russo2016nonclassical,james2019phase}.

Our results hinge on the development of a novel order parameter for local structure identification which is multidimensional and lossless, and is shown to successfully characterize these complex nucleation pathways and to identify local structures with high accuracy.
A proper polymorph decomposition, for example, is essential in the determination of the nucleation rate~\cite{cheng2018theoretical}.
We believe that the generality and flexibility of our method makes it suitable for the study of a large range of systems showing characteristic ordered or disordered signatures, such as defects or interfaces in crystalline or amorphous materials.

\begin{acknowledgments}
We acknowledge support from the European Research Council Grant DLV-759187. We thank A. Attanasi and M. Mosayebi for useful discussions.
\end{acknowledgments}


\appendix

\section{LOCAL STRUCTURE IDENTIFICATION METHODS}
\label{app:A}

\subsection{Common neighbor analysis (CNA)}

The Common neighbor Analysis (CNA) method \cite{honeycutt1987} assigns a structure type to every particle based on a nearest-neighbor graph accounting for the bond connectivity among neighbors of a given particle.
Particles are considered to be neighbors if they are closer to each other than a specific cutoff.
In the present work, for HS we employ the adaptive Common neighbor Analysis (a-CNA) method \cite{stukowski2012}, in which an optimal cutoff radius is automatically computed for each individual particle. 
A major disadvantage of CNA
is that no structure type is assigned to particles with unknown signatures, and it is sensitive to thermal fluctuations \cite{larsen2016}.

\subsection{Extended common neighbor analysis (Ext-CNA)}

In order to assign cubic or hexagonal diamond structure type to a water Oxygen, information on the position of his second nearest neighbors (i.e., second shell) are needed.
In the diamond structure, nearest neighbor Oxygens don't have common neighbors, and the second and third shells are not well separated.
In order to apply the CNA method to identify diamond structures, the extended CNA (Ext-CNA) has been introduced in Ref.\cite{maras2016}.
In the software \emph{Ovito} \cite{stukowski2010} it is available as \emph{Identify diamond structure} function.
In the Ext-CNA, the CNA method is applied to the 12 second nearest neighbors of a central particle, which are found as the first neighbors of the first 4 neighbors of the central particle under consideration. We refer to this CNA method to identify particles in the mW water model.
We also consider the method we name Ext-CNA-1st (available as option in \emph{Ovito}), which includes in the ice I structures also particles being first neighbors of a particle classified as ice I by the Ext-CNA method. These additional particles have four first neighbors positioned on the right lattice sites of the relative ice I structure, but at least one of its second nearest neighbors is off lattice.

\begin{figure*}[!ht] 
	\begin{center}
          \includegraphics[width=3.45cm]{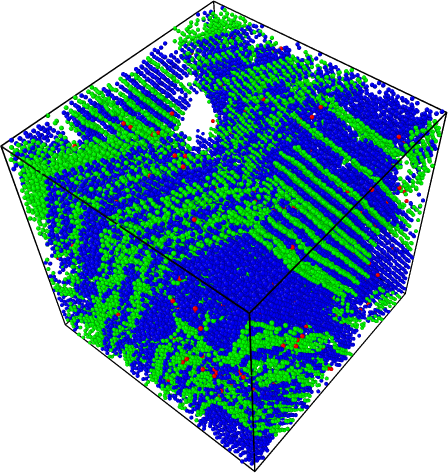}
          \includegraphics[width=3.45cm]{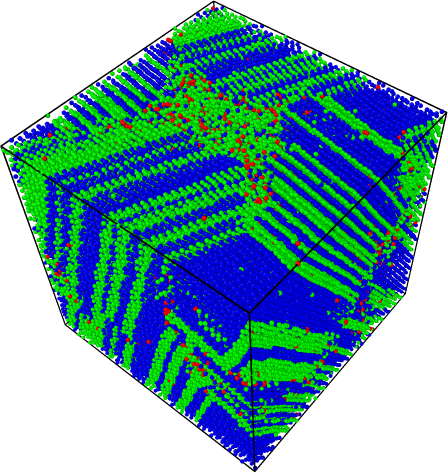}
          \includegraphics[width=3.45cm]{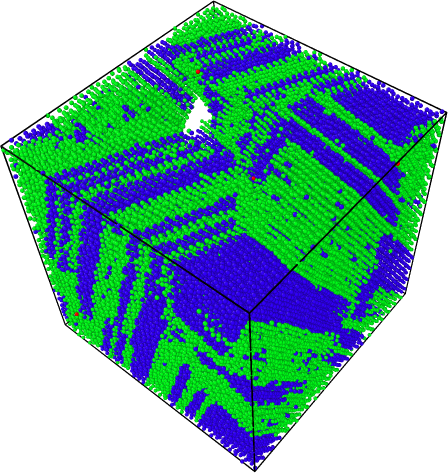}
          \includegraphics[width=3.45cm]{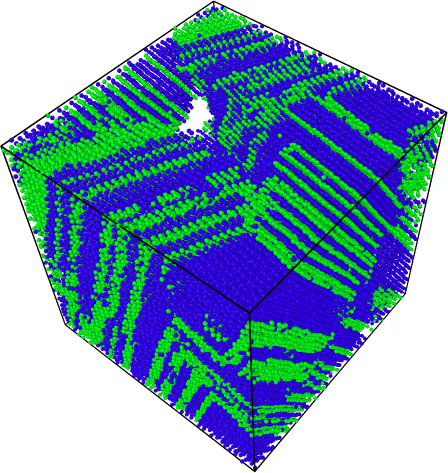}
          \includegraphics[width=3.45cm]{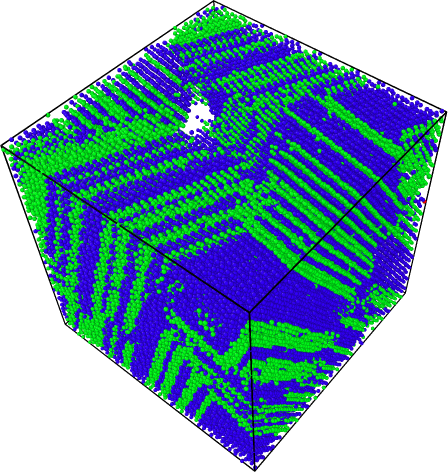}
            \caption{\label{fig:nc_HS2} Homogeneous nucleation of hard spheres. Particles structure from the same configuration (snapshot) have been identified using the following methods (from left to right): a-CNA, PTM, BOO up to second shell, BOO up to first shell, and LID.
    In all panels the colors associated to \emph{fcc}, \emph{hcp}, and \emph{bcc} structures are blue, green, and red, respectively. The calculation of a-CNA and PTM, and snapshots visualization have been obtained by using \emph{Ovito} \cite{stukowski2010}.}
        \end{center}
\end{figure*}

\subsection{Polyhedral template matching (PTM)}

This method is based on the topology of the local particle environment \cite{larsen2016}. 
It makes use of the convex hull formed by a fixed number of neighboring particles, which are identified using a Voronoi-based method. The planar graph representing the convex hull is used to classify structures.
PTM is less sensitive to thermal fluctuations respect to a-CNA, but it still requires the definition of reference structures.
In Fig.~\ref{fig:nc_HS2} we show the nucleus spanning the simulation box displayed in the main text in Fig.\ref{fig:nc_HS}, but here identified by using (from left to right) a-CNA, PTM, BOO up to second shell, BOO up to first shell, and LID.
We notice that BOO up to the second shell shows problems in distinguishing parallel layers of alternating phases when they are close to each other, while BOO up to the first shell improves the identification of those parallel layers of alternating phases, even though it gives similar results on the composition of polytypes respect to BOO up to second shell.

\subsection{Chill+}

Chill+ \cite{nguyen2015} classifies cubic ice, hexagonal ice, and clathrate hydrate structures in water. It is based on the identification of staggered and eclipsed bonds: since an Oxygen atom in crystalline ice is 4-coordinated (first neighboring shell), if we consider two neighboring Oxygen atoms, we can look at the cluster of 8 atoms composed by these two and their first neighbors.
Looking at the atoms along the axis of the bond between the first two atoms, if all the six neighboring atoms are visible we have a staggered bond, while if we see three neighboring atoms we have an eclipsed bond.
Because the presence of thermal fluctuations and other effects distorting bonds, as for other methods comparing local environments to a reference structure, thresholds to establish if a bond is close enough to the perfect staggered or eclipsed bond and then being identified with it have to be introduced.
In particular, if the bond order parameter $q_{3m}$ is between 0.25 and -0.35 the bond is eclipsed, while if it is less than -0.8 the bond is staggered.
The crystalline structure associated to an Oxygen atom depends on the number of eclipsed and staggered bonds. For example, hexagonal ice has 1 eclipsed and 3 staggered bonds, while cubic ice has all 4 bonds staggered.  
This method is specific for water.

\subsection{Bond orientational order (BOO)}

Steinhardt or bond orientational order (BOO) parameters $q_l(i)$ and $w_l(i)$ describe local order (as seen from particle i) in terms of spherical harmonics of order $l$. They are based on the following complex vector $q_{lm}(i)$ associated to the particle i
\begin{equation}
q_{lm}(i)=\dfrac{1}{N_b(i)}\sum_{j=1}^{N_b(i)}Y_{lm}(r_{lm})
\end{equation}
where $N_b(i)$ is the number of neighbors of particle i, $l$ is integer and $m$ is an integer running from $m=-l$ to $m=l$, $Y_{lm}(r_{ij})$ are the spherical harmonics and $r_{ij}$ is the position vector from particle $i$ to $j$; and on the averaged $\bar{q}_{lm}(i)$ defined as
\begin{equation}
\bar{q}_{lm}(i)=\dfrac{1}{N_b(i)+1}\sum_{k\in\{i,N_b(i)\}}q_{lm}(k)
\end{equation}
where the sum runs over the $N_b(i)$ plus the particle $i$.  
The local bond order, or Steinhardt, parameters $q_l(i)$ are defined as
\begin{equation}
q_l(i)=\sqrt{\dfrac{4\pi}{2l+1}\sum_{m=-l}^{l}|q_{lm}(i)|^2}.
\end{equation}
The parameter corresponding to a specific value of $l$ capture a specific crystal symmetry.
All $q_l(i)$ depend on the angles formed by neighboring particles and are independent of a reference frame.
The averaged Steinhardt OP $\bar{q}_l(i)$ are defined as
\begin{equation}
\bar{q}_l(i)=\sqrt{\dfrac{4\pi}{2l+1}\sum_{m=-l}^{l}|\bar{q}_{lm}(i)|^2}.
\end{equation}
Cubic Steinhardt OP $w_l(i)$ are defined as
\begin{equation}
\begin{array}{l}
w_l(i)=\\
\\
\dfrac{{\displaystyle\sum_{m_1+m_2+m_3=0}}\left(\begin{array}{ccc}l&l&l\\m_1&m_2&m_3\end{array}\right)q_{lm_1}(i)q_{lm_2}(i)q_{lm_3}(i)}{\left(\sum_{m=-l}^{l}|q_{lm}(i)|^2\right)^{3/2}}
\end{array}
\end{equation}
where the term in parentheses is the Wigner $3j$ symbol, while the cubic averaged Steinhardt OP $\bar{w}_l(i)$ are defined as
\begin{equation}
\begin{array}{l}
\bar{w}_l(i)=\\
\\
\dfrac{{\displaystyle\sum_{m_1+m_2+m_3=0}}\left(\begin{array}{ccc}l&l&l\\m_1&m_2&m_3\end{array}\right)\bar{q}_{lm_1}(i)\bar{q}_{lm_2}(i)\bar{q}_{lm_3}(i)}{\left(\sum_{m=-l}^{l}|\bar{q}_{lm}(i)|^2\right)^{3/2}}.
\end{array}
\end{equation}

Here we consider $\bar{q}_{12}$ only for the identification of the solid nucleus without distinguishing polytypes, which has been used in other works \cite{leoni2019,tanaka_review}, and two methods based on BOO OP for the identification of all phases: the standard $\bar{q}_4\bar{q}_6$ map, in which case the choice of the protocol to compute and partition the map can strongly affect its application; and a group of 30 BOO, as described in the following, such to considerably increase the dimensionality of the order parameter space which
allows to easily increase the separation between the different populations of local environments we want to discriminate between.
The OP we use as input for the Neural Networks (NN) is composed of the following 30 BOO: $q_l(i)$ with $l=3,4,...,12$, $\bar{q}_l(i)$ with $l=3,4,...,12$, $w_l(i)$ with $l=4,6,8,10,12$, and $\bar{w}_l(i)$ with $l=4,6,8,10,12$.
There are different ways to obtain first and second shells of neighbors in order to compute BOO, like SANN algorithm \cite{meel2012} or using a fixed cutoff (see Sec.\ref{sec:NC}). Here we consider the first neighbors shell as composed by the $N$ particles closer to the particle under investigation, and the second neighbors shell as composed by the $M$ particles closer to the particle under investigation, excluding the first $N$ particles. 
As $N$ and $M$ we consider $N=12$ and $M=6$ for HS, while $N=4$ and $M=12$ for mW water. These values for $N$ and $M$ are related to the number of first and second neighbors in the crystalline structures forming in these models.

\subsection{$\bar{q}_4\bar{q}_6$ sensitivity to protocols}

This method for local structure identification is very popular, but, as discussed in the main text, it is very sensitive to the way in which it is computed and to the thresholds used to partition the map.
Here we show that, when applied to the determination of the nucleus size and its composition of mW water, the $\bar{q}_4\bar{q}_6$ method can give very different results. 

First of all, in order to define the neighbors of a particle $i$, two approaches are usually employed: considering the $n_n$ particles closer to particle $i$ or considering all the $n_{cut}$ particles found at a distance from particle $i$ smaller than $r_{cut}$. 
Once the $\bar{q}_4\bar{q}_6$ map has been computed, it can be partitioned in different ways. 

\begin{figure}[!t] 
	\begin{center}
		\includegraphics[width=8.5cm]{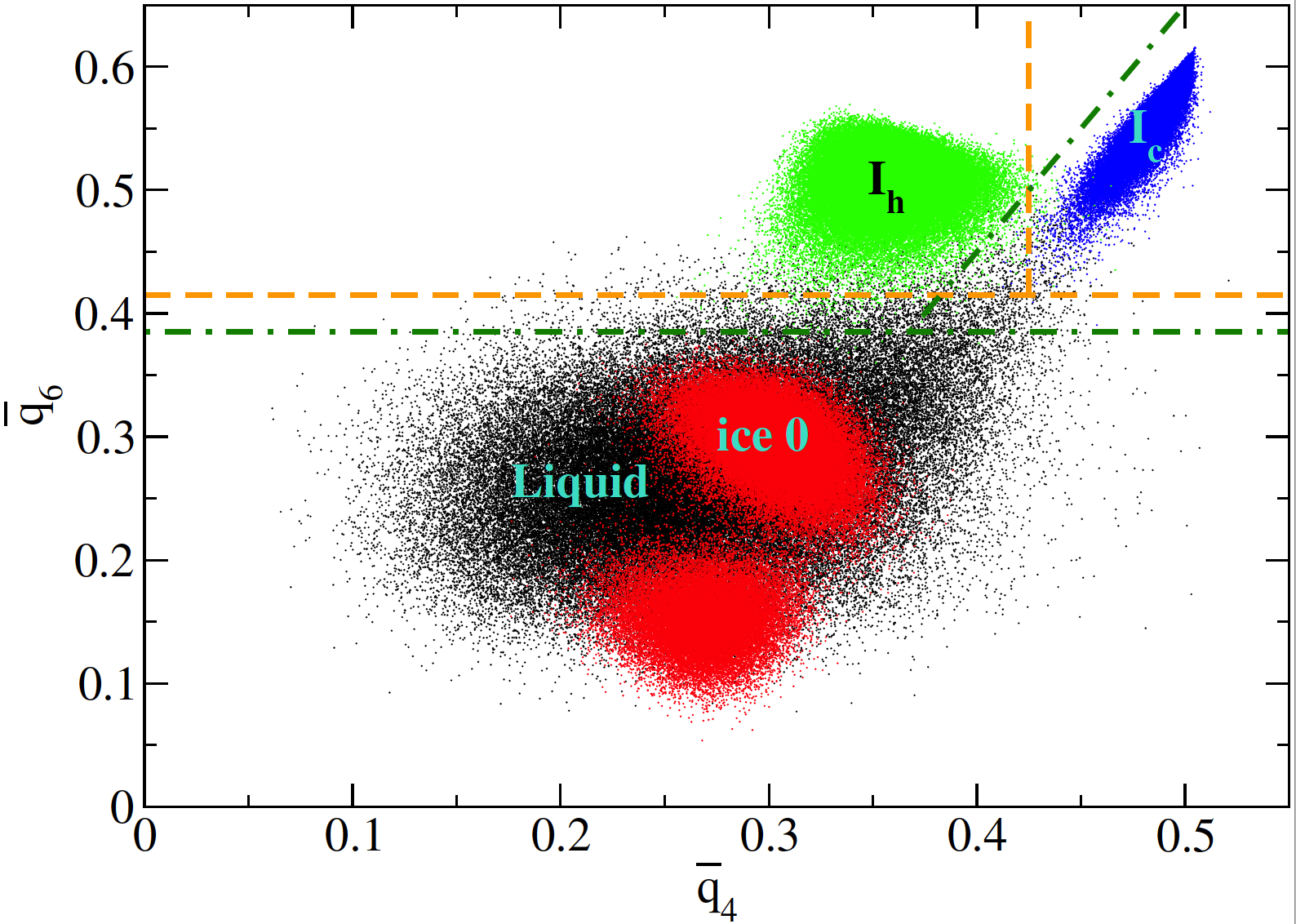}
		\caption{\label{fig:q4q6_LD-A} $\bar{q}_4\bar{q}_6$ map calculation and partition following method LD-A. Each dot corresponds to the $\bar{q}_4$, $\bar{q}_6$ coordinates associated to a particle of the following systems composed of N=5376 mW particles at melting: I$_c$ (blue), I$_h$ (green), ice~0 (red), and liquid water (black).}
	\end{center}
\end{figure}

In Fig.~\ref{fig:q4q6_LD-A} we show the $\bar{q}_4\bar{q}_6$ map obtained by 
considering $n_{cut}$ neighbors with $r_{cut}=1.43\sigma_{mW}$, and the particles phase is associated to fluid if $\bar{q}_6<0.415$, otherwise they are crystalline and in particular in the phase I$_c$ if $\bar{q}_4>0.425$, and I$_h$ otherwise (orange dashed lines correspond to these thresholds).
This method, named LD-A in Ref.~\cite{prestipino2018}, doesn't discriminate between the liquid phase and ice~0 (black and red dots corresponding to the fluid phase and ice~0, respectively, overlap and then cannot be distinguished, see Fig.~\ref{fig:q4q6_LD-A}).
%
\begin{figure}[!h] 
	\begin{center}
		\includegraphics[width=8.5cm]{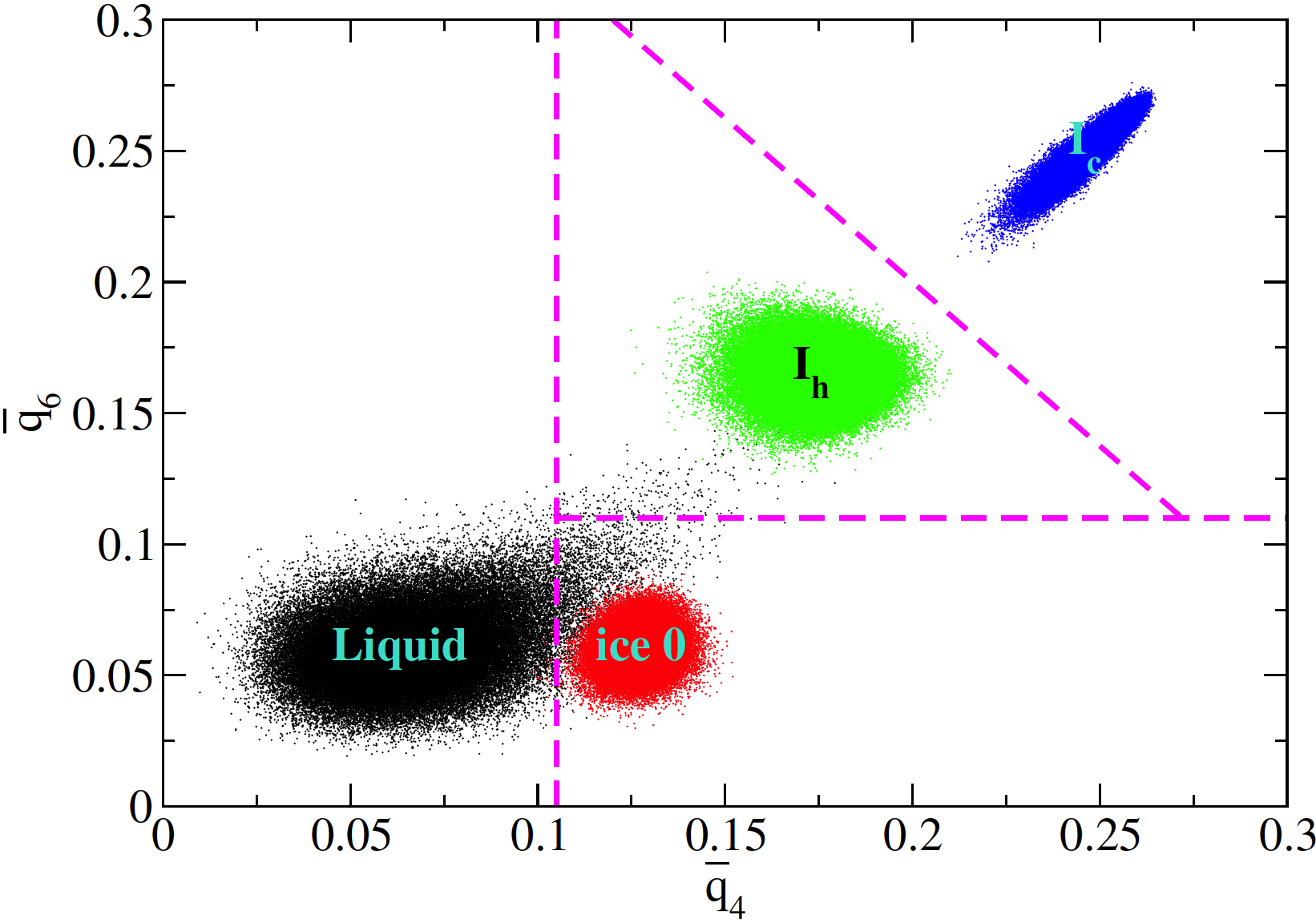}
		\caption{\label{fig:q4q6_LD-B} $\bar{q}_4\bar{q}_6$ map calculation and partition following method LD-A2.}
	\end{center}
\end{figure}

In Fig.~\ref{fig:q4q6_LD-A} we show also another choice of thresholds to partition 
the $\bar{q}_4\bar{q}_6$ map, that we name LD-A2: particles are fluid if $\bar{q}_6<0.385$, otherwise they are crystalline and in particular associated to the phase I$_c$ if $2\bar{q}_4>\bar{q}_6+0.35$, and I$_h$ otherwise (dark green dash-dotted lines correspond to these thresholds). This choice of thresholds allows to better partition the $\bar{q}_4\bar{q}_6$ map at melting (not shown here) respect to LD-A. 
\begin{figure}[!t] 
	\begin{center}
		\includegraphics[width=8.5cm]{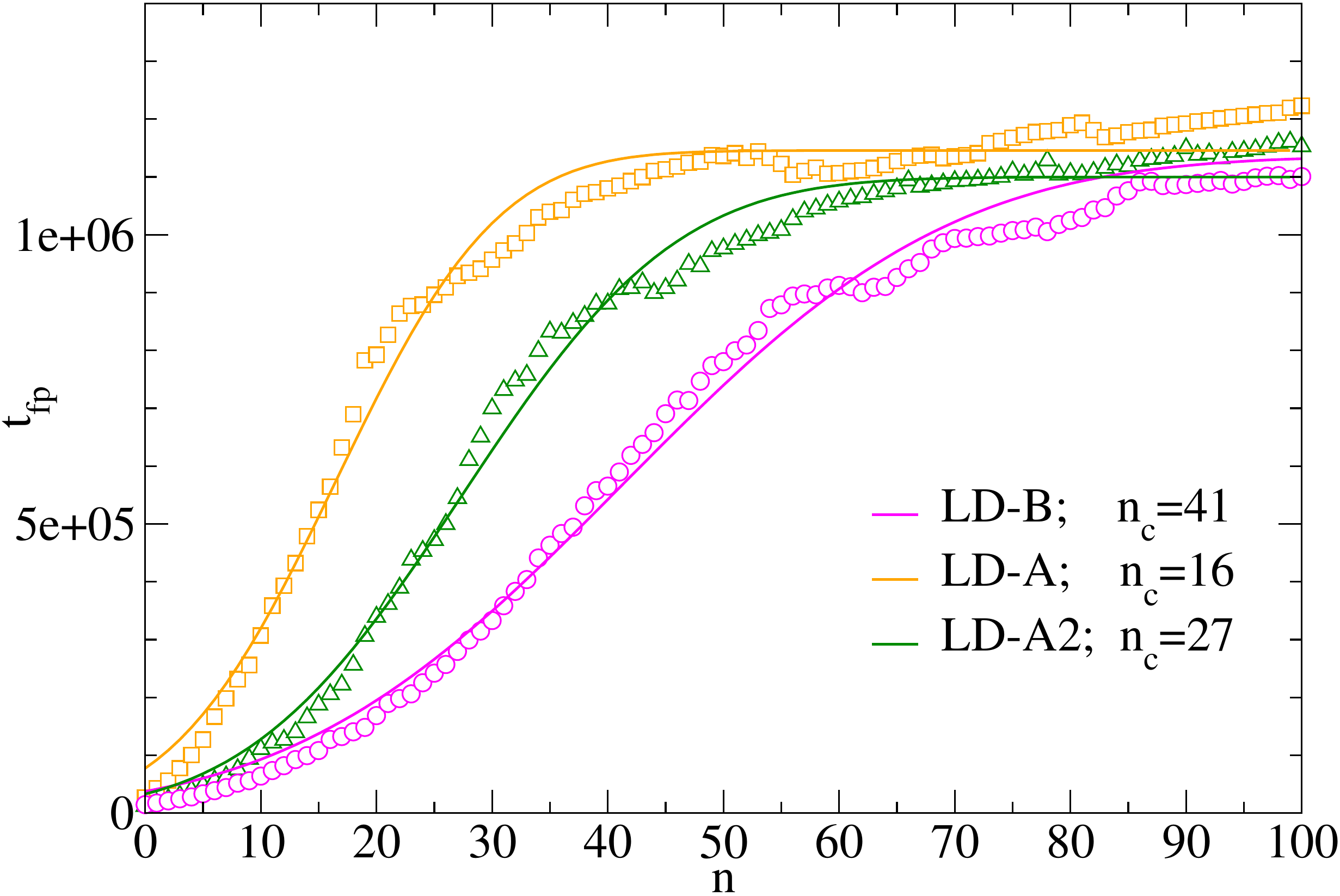}
        \caption{\label{fig:q4q6_FPT} Average first passage time $t_{f_p}$ as a function of nucleus size computed using the $\bar{q}_4\bar{q}_6$ methods LD-A (orange squares), LD-A2 (dark-green triangles), and LD-B (magenta circles).}
	\end{center}
\end{figure}
%

\begin{figure}[!t] 
	\begin{center}
		\includegraphics[width=8.5cm]{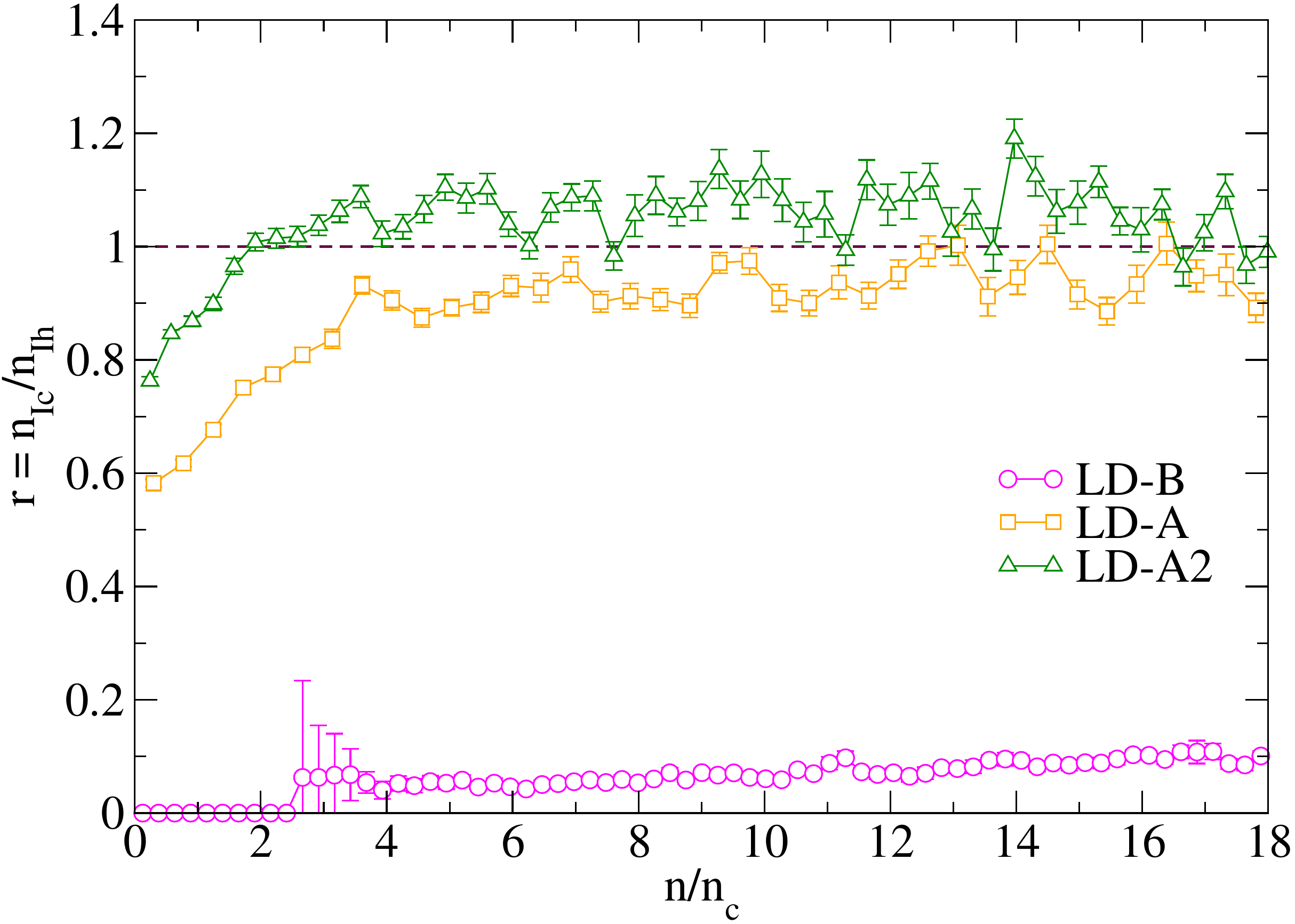}
		\caption{\label{fig:q4q6_r} Average ratio $r=n_{I_c}/n_{I_h}$ between the number of particles composing the nucleus in the cubic phase (I$_c$) and the hexagonal phase (I$_h$) using the $\bar{q}_4\bar{q}_6$ methods LD-A (orange squares), LD-A2 (dark-green triangles), and LD-B (magenta circles).
		$r$ is plotted against the nucleus size $n$ normalized by the critical nucleus size $n_c$ for each specific method.}
	\end{center}
\end{figure}

In Fig.~\ref{fig:q4q6_LD-B} we show another method to obtain the $\bar{q}_4\bar{q}_6$ map, where the number of neighbors is fixed to $n_n=16$,  and the threshold are the following: if $\bar{q}_4<0.105$ particles are fluid, while crystalline in the opposite case. Crystalline particles are classified as ice~0 if $\bar{q}_6<0.11$, and ice I in the opposite case. Ice I particles are classified as I$_c$ if $\bar{q}_4/0.36+\bar{q}_6/0.45>1$, and I$_h$ otherwise.
This method, named LD-B in Ref.~\cite{prestipino2018}, allows to discriminate between the liquid phase and ice~0.
In all cases the $\bar{q}_4\bar{q}_6$ map is computed at the nucleation temperature $T=204$~K and pressure $P=0$~Pa.

In Fig.~\ref{fig:q4q6_FPT} we show the average first passage time $t_{fp}$, described in Sec.~\ref{sec:critical_nucleus}, as a function of the nucleus size $n$ obtained applying the three different methods considered here to compute and partition the $\bar{q}_4\bar{q}_6$ map. We can notice the big variation in the value of the critical nucleus size $n_c$ estimated from the different methods.

In Fig.~\ref{fig:q4q6_r} we show the ratio $r$ between the number of particles $n_{I_c}$ in the cubic phase and the number of particles $n_{I_h}$ in the hexagonal phase found in the nucleus as a function of its size $n$ divided by the critical nucleus size $n_c$ applying the three different methods considered here to get the $\bar{q}_4\bar{q}_6$ map.
As found for the average first passage time, also in this case each method gives a different estimation of $r$ (averaging only on the stationary part, that is excluding small cluster size): 0.94, 1.07, and 0.07 for LD-A, LD-A2, and LD-B, respectively. 
Even though LD-B is able to discriminate between the liquid phase and ice~0, it is strongly biased towards the hexagonal phase.

\begin{figure*}[!t] 
	\begin{center}
		\includegraphics[width=17cm]{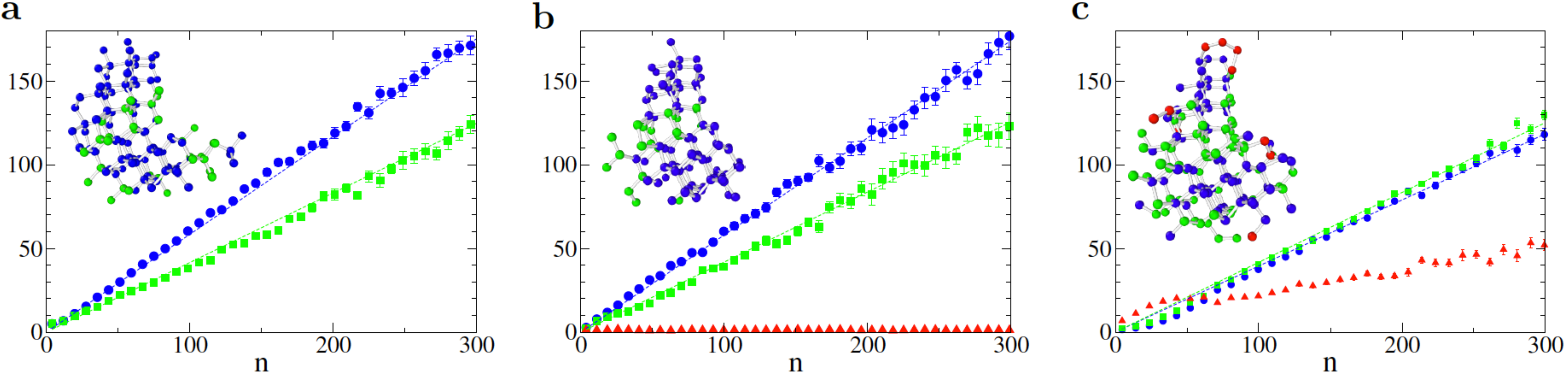}
                \caption{\label{fig:nc_Water_A} Homogeneous nucleation of mW water: average composition of the main cluster as identified by (a) CNA up to first neighbors, (b) BOO, and (c) LID. Insets in (a), (b) and (c) show a typical nucleus composed of 179 (66 if disregarding I$_c$ and I$_h$ first neighbors), 104 and 144 particles, respectively.
    In all panels the colors associated to I$_c$, I$_h$, ice~0 structures are blue, green, and red, respectively.}
        \end{center}
\end{figure*}

\begin{table*}[!ht]
\centering
\begin{tabular}{|c|c||c|c|c||c|c|c||c|c|c||c|c|c||c|c|c|}
\hline
\multicolumn{2}{|c||}{\bf{TEST I$_c$}} &
\multicolumn{3}{|c||}{\bf{Ext-CNA}} &
\multicolumn{3}{|c||}{\bf{Ext-CNA-1st}} &
\multicolumn{3}{|c||}{\bf{Chill+}} &
\multicolumn{3}{|c||}{\bf{BOO}} &
\multicolumn{3}{|c|}{\bf{LID}}\\
\hline
Region & Particles & I$_c$ & I$_h$ & L & I$_c$ & I$_h$ & L & I$_c$ & I$_h$ & L & I$_c$ & I$_h$ & L & I$_c$ & I$_h$ & L\\
\hline
1 & 16.4 & 96.78 & 0.00 & 3.22 & 98.43 & 1.27 & 0.30 & 97.10 & 0.00 & 2.90 & 93.86 & 3.78 & 2.36 & 99.54 & 0.18 & 0.28\\
\hline
2 & 7.25 & 10.28 & 0.00 & 89.72 & 11.73 & 73.50 & 14.77 & 75.56 & 0.14 & 24.30 & 64.15 & 17.58 & 18.27 & 67.52 & 1.55 & 30.93\\
\hline
3 & 7.8 & 0.60 & 0.00 & 99.40 & 34.22 & 45.58 & 20.20 & 57.76 & 1.53 & 40.71 & 63.51 & 9.52 & 26.97 & 10.93 & 1.58 & 87.49\\
\hline
4 & 9.05 & 0.12 & 0.00 & 99.88 & 16.37 & 60.04 & 23.60 & 17.69 & 1.18 & 76.13 & 30.17 & 5.48 & 59.35 & 4.15 & 0.72 & 95.08\\
\hline
\end{tabular}
\caption{Benchmark of different methods for the identification of particle in the I$_c$ phase present in nuclei of size $n=200$. L refers to the liquid phase. Except the number indicating the region, all the other numbers refer to percentages.}
\label{tab:Ic}
\end{table*}
%
\begin{table*}[!ht]
\centering
\begin{tabular}{|c|c||c|c|c||c|c|c||c|c|c||c|c|c||c|c|c|}
\hline
\multicolumn{2}{|c||}{\bf{TEST I$_h$}} &
\multicolumn{3}{|c||}{\bf{Ext-CNA}} &
\multicolumn{3}{|c||}{\bf{Ext-CNA-1st}} &
\multicolumn{3}{|c||}{\bf{Chill+}} &
\multicolumn{3}{|c||}{\bf{BOO}} &
\multicolumn{3}{|c|}{\bf{LID}}\\
\hline
Region & Particles & I$_c$ & I$_h$ & L & I$_c$ & I$_h$ & L & I$_c$ & I$_h$ & L & I$_c$ & I$_h$ & L & I$_c$ & I$_h$ & L\\
\hline
1 & 16.4 & 0.00 & 96.01 & 3.99 & 3.06 & 96.37 & 0.58 & 0.00 & 97.11 & 2.89 & 4.53 & 93.63 & 1.84 & 0.51 & 98.85 & 0.64\\
\hline
2 & 7.25 & 0.00 & 4.94 & 95.06 & 68.31 & 24.26 & 7.43 & 0.00 & 76.70 & 23.30 & 19.63 & 73.40 & 6.97 & 5.98 & 38.84 & 55.18\\
\hline
3 & 7.8 & 0.00 & 1.65 & 98.35 & 71.32 & 10.76 & 17.92 & 0.00 & 64.50 & 35.50 & 11.43 & 65.48 & 23.09 & 0.64 & 8.75 & 90.61\\
\hline
4 & 9.05 & 0.00 & 0.00 & 100.00 & 41.29 & 21.38 & 37.33 & 1.67 & 9.69 & 78.64 & 19.68 & 57.66 & 22.66 & 0.41 & 2.27 & 97.32\\
\hline
\end{tabular}
\caption{Benchmark of different methods for the identification of particle in the I$_h$ phase present in nuclei of size $n=200$. L refers to the liquid phase. Except the number indicating the region, all the other numbers refer to percentages.}
\label{tab:Ih}
\end{table*}

\subsection{Composition of mW nuclei}

Similarly to Fig.~\ref{fig:nc_HS}, we show in Fig.~\ref{fig:nc_Water_A} the average fractional composition as a function of nucleus size for mW molecules at ambient pressure and temperature $T=204$~K as identified by EXT-CNA-1st (panel \textit{a}), BOO (panel \textit{b}), and LID (panel \textit{c}).

\subsection{Benchmark}\label{sec:benchmark}

Considering the wide variation of results on the nucleus properties predicted by
different methods adopted in the literature, some of which analyzed here, it would be desirable to find benchmarks to evaluate accuracy and reliability of each of them.
Here we propose a simple test in which we know by construction the phase of each particle belonging to the nucleus and we use different methods to identify them.
We consider a cluster composed of particles of both phases ice I$_c$ and I$_h$ 
obtained from a perfect lattice of stacking ice with alternated layers of I$_c$ and I$_h$ 
at a density $\rho=0.982$~g/cm$^3$ corresponding to the temperature $T=235$K at equilibrium conditions (see Ref.~\cite{leoni2019}).
We obtain a cluster of size $n=200$ following the minimum energy rule described in Ref.~\cite{leoni2019}.
Then we let the cluster equilibrate in contact with a liquid phase of density $\rho=1.002$~g/cm$^3$, corresponding to equilibrium conditions at the temperature $T=235$K, using fixed-topology MC simulations (see Ref.~\cite{leoni2019}) which allow for bonds elongation up to a maximum cutoff (set to 1.3~\AA), while keeping the topology fixed.

Since we know the phase (I$_c$ or I$_h$) of each particle composing the cluster, using different methods we identify each particle phase and compare this prediction with its true value.
We distinguish particles of the cluster as belonging to different regions depending on the number of theirs first neighbors ($fn$) and the sum of first neighbors of first neighbors ($fn2$) in the following way: for all regions under consideration $fn=4$, while $fn2=16, 15, 14, 13$ for regions 1,2,3, and 4, respectively. Only particles belonging to region 1 have a fully formed second shell.

In Table~\ref{tab:Ic} we show (second column) the average percentage of particles of the cluster belonging to each region (first column), and the percentage of particles correctly identified as I$_c$, or incorrectly identified as I$_h$ or as liquid phase, L, for the different methods (columns from third to seventeenth).
In Table~\ref{tab:Ih} we show the same results, but for the identification of I$_h$.
For example, for clusters of size $n=200$ considered here, particles belonging to region 1 are in average only the $16.4\%$ of the total.
These results are obtained by averaging over 20 different clusters realized by using the minimum energy rule and 10 different evolution times.
The identification method $\bar{q}_4\bar{q}_6$ is strongly affected by the choice of the protocol used to compute and partition it (see Ref.~\cite{prestipino2018} and Appendix \ref{app:A}), and then it is not shown in the tables.

From Tabs.~\ref{tab:Ic},\ref{tab:Ih} we can see for example that the method Ext-CNA identifies correctly the $\sim$96\% of times cubic and hexagonal ice particles in region 1. When considering other regions the percentage of correct identification quickly goes to zero for increasing region label, that is for more and more incomplete second shells, in which case particles are more likely associated to a liquid phase.  
This is reflected in the very small value of the critical cluster size, $n_c=4$, obtained with this method (see Fig.~\ref{fig:FPT}).
In the case of Ext-CNA-1st the performance in the region 1 is similar to the method Ext-CNA, while particles in other regions are mainly identified as crystalline.
However, as we also noted by snapshots inspection, 
in regions 2,3 and 4 Ext-CNA-1st misidentifies crystalline particles, often associating I$_c$ phase to I$_h$ particles and vice versa. 
This is not surprising considering that Ext-CNA-1st associates to the first neighbors of a particle in the I$_c$ (I$_h$) phase the same I$_c$ (I$_h$) phase (see Appendix \ref{app:A}), and nuclei tested in the present benchmark are composed of alternating layers of I$_c$ and I$_h$ phases.
For this reason, when using the {\it Identify diamond structure} function of Ovito, it would be important to specify if also first neighbors or even second neighbors of crystalline particles are included in the method to compute quantities like for example the cubicity which gives a measure of the amount of I$_c$ respect to I$_h$ composing the nucleus. 
Finally, BOO shows a good identification rate with limited misidentifications, while LID and Chill+ give the best performance with extremely low misidentifications.


A conservative way to rate the performance of a method from  
these benchmarks is to evaluate the percentage of particles correctly identified in region 1 (particles with fully formed second shell), and considering the relevance of misidentification decreasing for increasing region label.
From these considerations we conclude that Ext-CNA is too conservative missing many crystalline particles of the nucleus, while Ext-CNA-1st is affected by an important misidentification of crystalline particles with incomplete second shells.
BOO shows low misidentification of crystalline particles.
On the other hand, LID and Chill+ are the two methods with the lowest misidentification, with LID showing the best performance for identification of crystalline phases in region 1.

In order to evaluate the influence of thermal fluctuations on the particle identification methods, we repeated the previous benchmark, but this time considering rigid clusters (no bonds elongation) equilibrated with the liquid phase.
Also in this case we observe a similar behavior of the different methods.

%
\subsection{Correlation between precursors and OP}

\begin{figure}[!b] 
	\begin{center}
		\includegraphics[width=3.8cm]{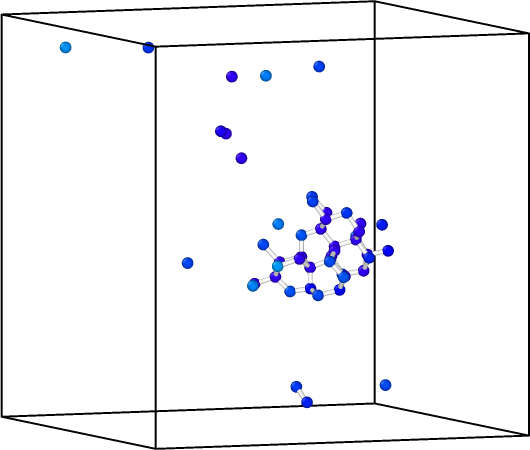}
		\includegraphics[width=3.8cm]{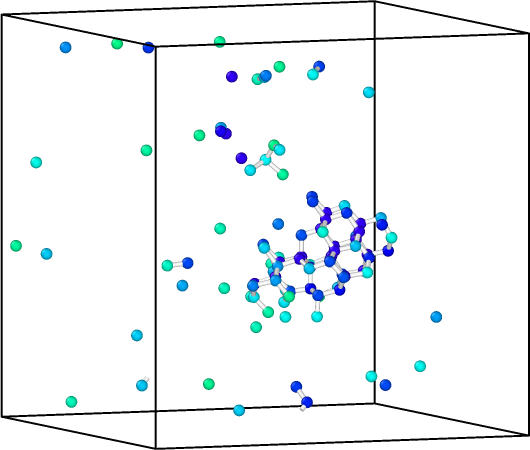}
		\includegraphics[width=0.35cm]{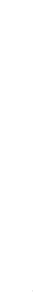}

		\includegraphics[width=3.8cm]{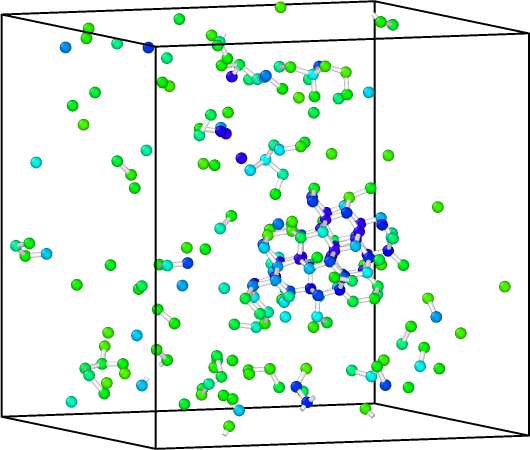}
		\includegraphics[width=3.8cm]{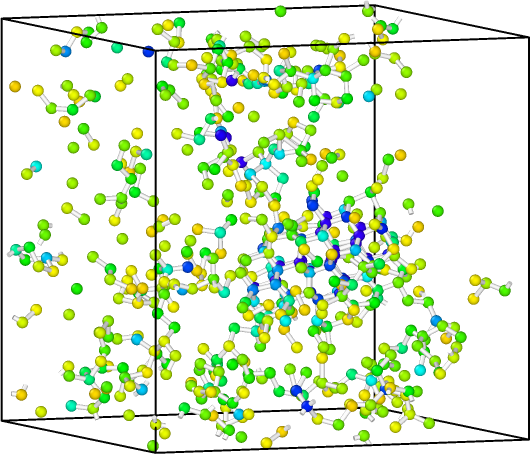}
		\includegraphics[width=0.35cm]{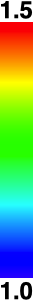}

				\caption{\label{fig:snap_th} Snapshots of a nucleation trajectory at time $t=102$ in $10^4$ MC steps units to which it corresponds the presence of a nucleus identified with LID of size $n=70$ (see Fig.~\ref{fig:N_vs_t}). Colors are assigned to particles which distance $d_i^{LID}$ is smaller than 1.1 (top-left panel) 1.2 (top-right panel) 1.3 (bottom-left panel) or 1.4 (bottom-right panel).}
	\end{center}
\end{figure}
%
\begin{figure}[!h]
	\begin{center}
		\includegraphics[width=3.8cm]{pos_102_th11.png}
		\includegraphics[width=3.8cm]{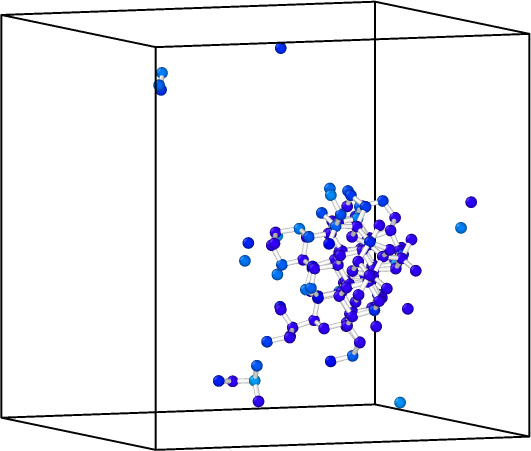}

		\includegraphics[width=3.8cm]{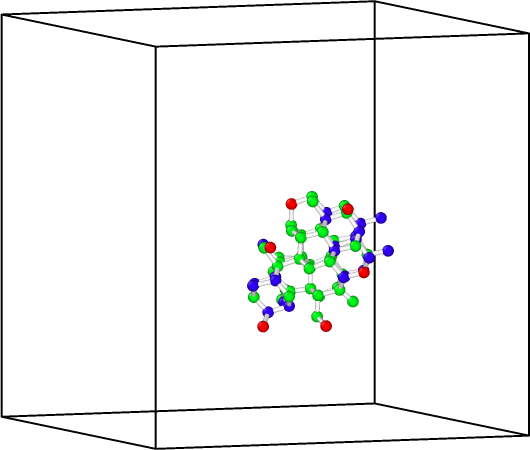}
		\includegraphics[width=3.8cm]{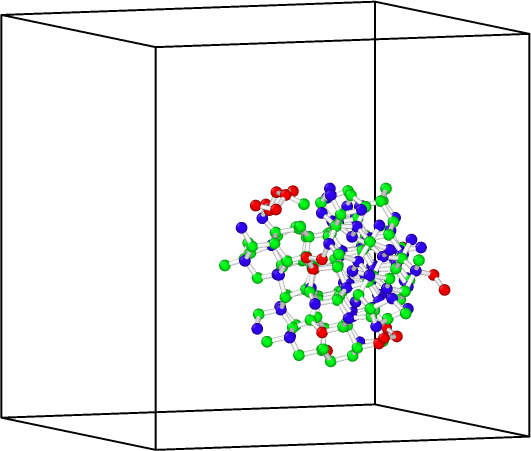}
		
		\includegraphics[width=3.8cm]{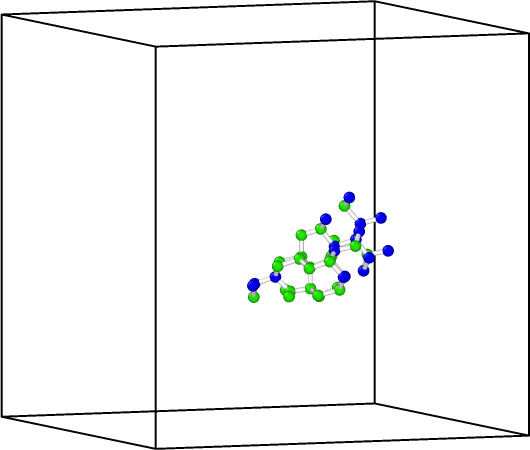}
		\includegraphics[width=3.8cm]{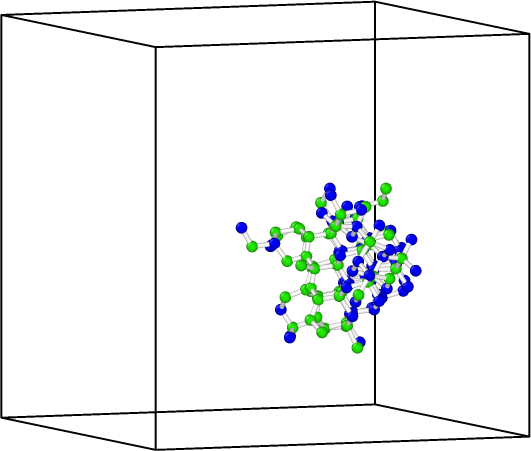}

		\caption{\label{fig:snap_dij-C+} Snapshots of a nucleation trajectory at time $t=102$ (left panels) and $t=141$ (right panels) in $10^4$ MC steps units showing the presence of a nucleus identified with:
		(upper panels) particles which distance $d_i^{LID}$ is smaller than 1.1; (middle panels) LID (Blue, green, and red for I$_c$, I$_h$, and ice~0, respectively) giving a nucleus of size $n=70$ (left panel) and $n=197$ (right panel) (see Fig.~\ref{fig:N_vs_t}); (lower panels) Chill+ (Blue and green for I$_c$ and I$_h$, respectively).}
	\end{center}
\end{figure}

For each particle $i$ we compute the Euclidean distance $d_i^{LID}$ between the vector LID at a specific time and the LID associated to the perfect crystalline structure. Here we consider as reference the LID signal associated to I$_c$, as very similar results are obtained with respect to I$_h$ (not shown).
In the following we show the value of $d_i^{LID}$ associated to each particle of a sample at two specific times (see Fig.~\ref{fig:N_vs_t}) at which the nucleus has a size of $n=70$ (Fig.~\ref{fig:snap_th},\ref{fig:snap_dij-C+}) and $n=197$ (Fig.~\ref{fig:snap_dij-C+}), (see the green square for $n=70$ and the violet circle for $n=197$ in Fig.~\ref{fig:N_vs_t}).
In Fig.~\ref{fig:snap_th} we show snapshots corresponding to the nucleus of size $n=70$, where particles $i$ with a distance $d_i^{LID}$ smaller than 1.1, 1.2, 1.3, and 1.4 (from left to right and from top to bottom) are shown with a color code going from 1.0 (blue) to 1.5 (red).
The field $d_i^{LID}$ correlates with crystalline structures present in the system (see top-left snapshot in Fig.~\ref{fig:snap_th}), and in particular with the main cluster as detected by other methods (see Fig.~\ref{fig:snap_dij-C+}).

In Fig.~\ref{fig:snap_dij-C+}, from top to bottom, we show: particles with the Euclidean distance $d_i^{LID}<1.1$ (see Fig.~\ref{fig:snap_th} for color codes); particles belonging to the main cluster as identified by using LID (blue for I$_c$, green for I$_h$, and red for ice~0); and particles belonging to the main cluster as identified by using the Chill+ algorithm (blue for I$_c$, and green for I$_h$).
Left (right) column in Fig.~\ref{fig:snap_dij-C+} refers to a snapshot of the nucleation trajectory shown in Fig.~\ref{fig:N_vs_t} at the time $t=102$ ($t=141$) in $10^4$ MC steps units. 
From Fig.~\ref{fig:snap_dij-C+} we can see that $d_i^{LID}$ correlates very well with the nucleus identified by LID and Chill+, and that the latter method, apart from not providing ice~0 particles, finds a smaller nucleus, as expected from its ability to estimate a smaller value of the critical nucleus respect to LID (see Fig.~\ref{fig:FPT}).

\newpage

\bibliography{biblio}

\end{document}